\documentclass[a4paper,12pt]{article}
\pdfoutput=1 

\usepackage{jcappub} 
\usepackage{subfigure}
\usepackage[T1]{fontenc} 

\def\hmath$#1${\texorpdfstring{{\rmfamily\textit{#1}}}{#1}}
\usepackage{hyperref}
\usepackage{amsfonts}
\usepackage{threeparttable}
\newcommand{\MEG}{\mu\rightarrow e\gamma}

\subheader{IIPDM-2019}

\title{\boldmath Lepton Flavor Violation and Collider Searches in a Type I + II Seesaw Model}

\author[a]{Manoel M. Ferreira,}
\author[b,c]{Tessio B. de Melo,}
\author[d]{Sergey Kovalenko,}
\author[e]{Paulo R. D. Pinheiro,}
\author[b]{Farinaldo S. Queiroz}

\affiliation[a]{Departamento de Fisica, Universidade Federal do Maranhao,
Campus Universitario do Bacanga, Sao Luis - MA, 65080-805 - Brazil}
\affiliation[b]{International Institute of Physics, Universidade Federal do Rio Grande do Norte,
Campus Universitario, Lagoa Nova, Natal-RN 59078-970, Brazil}
\affiliation[c]{Departamento de Fisica, Universidade Federal da Paraiba,
Caixa Postal 5008, 58051-970, Joao Pessoa-PB, Brazil}
\affiliation[d]{Universidad Tecnica Federico Santa Maria and Centro Cientifico-Tecnologico de Valparaiso
Casilla 110-V, Valparaiso, Chile}
\affiliation[e]{Coordenadoria Interdisciplinar de Ci\^encia e Tecnologia, 
Universidade Federal do Maranh\~ao,
65080-805, S\~ao Lu\'is - Maranh\~ao, Brazil}

\emailAdd{farinaldo.queiroz@iip.ufrn.br}

\abstract{
Neutrino are massless in the Standard Model. The most popular mechanism to generate neutrino masses are the type I and type II seesaw, where right-handed neutrinos and a scalar triplet are augmented to the Standard Model, respectively. In this work, we discuss a model where a type I + II seesaw mechanism naturally arises via spontaneous symmetry breaking of an enlarged gauge group. Lepton flavor violation is a common feature in such setup and for this reason, we compute the model contribution to the $\mu \rightarrow e\gamma$ and $\mu \rightarrow 3e$ decays. Moreover, we explore the connection between the neutrino mass ordering and lepton flavor violation in perspective with the LHC, HL-LHC and HE-LHC sensitivities to the doubly charged scalar stemming from the Higgs triplet. Our results explicitly show the importance of searching for signs of lepton flavor violation in collider and muon decays. The conclusion about which probe yields stronger bounds depends strongly on the mass ordering adopted, the absolute neutrino masses and which much decay one considers. In the $1-5$~TeV mass region of the doubly charged scalar, lepton flavor violation experiments and colliders offer orthogonal and complementary probes. Thus if a signal is observed in one of the two new physics searches, the other will be able to assess whether it stems from a seesaw framework.
}

\begin{document}

\maketitle
\flushbottom
\newpage

\section{Introduction}

The observation of neutrino oscillations implies in non-zero neutrino masses. Their masses are much smaller than any other in the Standard Model (SM) spectrum. Several mechanism have surfaced trying to explain the smallness of the neutrino masses \cite{Minkowski:1977sc,Mohapatra:1979ia,Yanagida:1980xy,Schechter:1980gr,Schechter:1981cv}. In the type I seesaw, the existence of heavy right-handed neutrinos is evoked, whereas in the type II seesaw a scalar triplet is added to the SM.  This scalar triplet couples to the SM lepton doublets and features a neutral scalar that develops a small vacuum expectation value giving rise to tiny neutrino masses  \cite{Mohapatra:1999zr,Montero:2000rh,Rodejohann:2004cg,Akhmedov:2006de,Gogoladze:2008gf}. This mechanism can elegantly explain the masses of the three active neutrinos in the SM and lead to several phenomenological imprints in collider, low energy observables and leptogenesis \cite{Sahu:2004ny,Sarma:2006xk,Sarma:2006xk,Melfo:2011nx,CarcamoHernandez:2017cwi,Chan:2007ng,Perez:2008ha,Akhmedov:2008tb}.\\

On the other side, we do not know why we have three generations of fermions in the SM. Theoretically speaking, it would nice to have a model where these two problems are simultaneously addressed. Several models have been proposed where the number of generations is addressed by anomaly cancellation and asymptotic freedom arguments, and they are known as 3-3-1 models \cite{Pisano:1991ee,Foot:1992rh,Foot:1994ym}. These models have an enlarged gauge sector, $SU(3)_c \times SU(3)_L \times U(1)_N$\, (3-3-1 for short). Notice that $SU(2)_L\times U(1)_Y$ in the SM gives place to $SU(3)_L \times U(1)_N$. Therefore, the fermions will now be arranged in the fundamental representation of $SU(3)_L$, i.e. triplets. The same happens for scalars fields, also sorted in triplets in order to generate fermion masses. Several versions based on the 3-3-1 symmetry have been proposed in the literature where the replication of fermion generations are explained \cite{Dias:2005xj,Dias:2006ns,Dias:2009au,Benavides:2010zw,Rodriguez:2016cgr}.\\

Not all of these realizations are capable of explaining neutrino masses at the same time, though. In this work, we focus on a model which can account for neutrino masses, known as {\it 3-3-1 model with right-handed neutrinos} \cite{Foot:1994ym,Hoang:1995vq,Hoang:1996gi,Hoang:1999yv}. Originally the model has only three scalar triplets. With scalar triplets one can generate two mass degenerate neutrinos and a massless one \cite{Chang:2006aa}, which is in conflict with existing data \cite{Tanabashi:2018oca}. The use of high dimensional effective operators has been put forth in the attempt to break the degeneracy \cite{Dias:2005yh,Queiroz:2010rj}. In this setup, the smallness of the neutrino masses and the oscillation pattern is not successfully explained. The most simple way to nicely solve this issue is by adding a scalar sextet \cite{Diaz:2003dk,Cogollo:2009yi}. The interesting aspect of this scalar sextet is that after spontaneous symmetry breaking it breaks down to a scalar triplet, two doublet scalars and a scalar singlet field. The scalar triplet is exactly the one desired to perform the type II seesaw mechanism. Therefore, in this way, the type II seesaw arises as a result of the spontaneous symmetry breaking. Several phenomenological aspects of the {\it 3-3-1 with right-handed neutrinos} have been explored in the past \cite{Gutierrez:2004sba,CarcamoHernandez:2005ka,Cogollo:2007qx,deS.Pires:2007gi,RamirezBarreto:2007cie,Cogollo:2008zc}, but our work differs from those because,

\begin{itemize}

\item We explicitly compute the $\MEG$ and $\mu\rightarrow 3e$ decays;

\item We explore their connection to neutrino mass ordering;

\item We put our results into perspective with collider and lepton flavor violation bounds. 
\end{itemize}

This work is structured as follows: In section II we briefly describe the model and explain how neutrino masses are incorporated, in section III we present the relevant collider bounds, in section IV we derive the lepton flavor violating muon decays and then draw our conclusions. In {\it Appendix}, we provide a details discussion of our reasoning and a general derivation of the $\MEG$ decay.  

\section{Model with right-handed neutrinos and scalar sextet}


There are several models based on the 3-3-1 gauge group. The model discussed here differs from the original proposal \cite{Pisano:1991ee,Foot:1992rh,Foot:1994ym} because of the presence of three right-handed neutrinos. This model does not suffer from a Landau pole at the TeV scale as the previous version. For this reason, it has attracted lots of attention in the past decade \cite{CiezaMontalvo:2006zt,Benavides:2009cn,Mizukoshi:2010ky,Alvares:2012qv,Cogollo:2012ek,Coutinho:2013lta,Kelso:2013nwa,Hernandez:2013hea,Dong:2014wsa,Salazar:2015gxa,Queiroz:2016gif,Hernandez:2016eod}. In this section, we will briefly discuss the model to ease our reasoning. We start with the fermion content.

\subsection{Fermion Content}


In this model, the left-handed leptons are arranged in the fundamental representation of $SU(3)_L$, whereas right-handed leptons as a singlet, as follows,  
\begin{equation}
\psi _{aL}=\left(
\begin{array}{c}
\nu _{a} \\
e_{a} \\
\nu _{a}^{c}%
\end{array}%
\right) _{L}\sim \left( \mathbf{1},\mathbf{3},-1/3\right) ,\text{ \ }%
e_{aR}\sim (\mathbf{1},\mathbf{1},-1),
\end{equation}%
where $a=1,2,3$ runs through the fermion generations. The numbers in parenthesis represent the quantum number of the fields under $SU(3)_C$, $SU(3)_L$ and $U(1)_N$, respectively. For instance, the $\psi_{aL}$ is a singlet under $SU(3)_C$, triplet under $SU(3)_L$ and has a charge equal $-1/3$ under $U(1)_N$.  Notice that the third component of the leptonic triplet is a right-handed neutrino since $\left(
\nu _{a}^{c}\right) _{L}=\nu _{R}^{c}$, explaining why this model is known as {\it 3-3-1 model with right-handed neutrinos}.\\

In a similar vein, the quarks are arranged as follows,

\begin{eqnarray}
Q_{\alpha L} &=&\left(
\begin{array}{c}
d_{\alpha } \\
-u_{\alpha } \\
D_{\alpha }%
\end{array}%
\right) _{L}\sim \left( \mathbf{3},\mathbf{3}^{\ast },0\right) ,\text{ } \\
u_{\alpha R} &\sim &\left( \mathbf{3},\mathbf{1},\frac{2}{3}\right) ,\text{ }%
d_{\alpha R}\sim \left( \mathbf{3},\mathbf{1},-\frac{1}{3}\right) ,\text{ }%
D_{\alpha R}\sim \left( \mathbf{3},\mathbf{1},-\frac{1}{3}\right) .
\end{eqnarray}%

\begin{eqnarray}
Q_{3L} &=&\left(
\begin{array}{c}
u_{3} \\
d_{3} \\
U%
\end{array}%
\right) _{L}\sim \left( \mathbf{3},\mathbf{3},1/3\right) ,\text{ } \\
u_{3R} &\sim &\left( \mathbf{3},\mathbf{1},\frac{2}{3}\right) ,\text{ }%
d_{3R}\sim \left( \mathbf{3},\mathbf{1},-\frac{1}{3}\right) ,\text{ }%
U_{R}\sim \left( \mathbf{3},\mathbf{1},\frac{2}{3}\right) .
\end{eqnarray}where $\alpha =1,2$. We highlight that the first two generations of quarks are placed in the anti-triplet representation of $SU(3)_L$. This has to be the case in order to cancel gauge anomalies. The quarks $D_{1,2}$
are down-type heavy quarks, in the sense, they have the same hypercharge as the SM quark down. A similar logic applies to the heavy up-quark $U$. These quarks have masses proportional to the scale of symmetry breaking of the 3-3-1 symmetry which should lie at several TeV to be consistent with the non-observation of such exotic quarks \cite{Coutinho:1999hf,Serkin:2019jiv}. The scale of symmetry breaking of the 3-3-1 symmetry will be kept sufficiently high to be consistent with this result. We will address the fermion masses below.

\subsection{Fermion Masses}

To generate fermion masses we need to invoke three scalar triplets of the following form, 

\begin{equation}
\chi =\left(
\begin{array}{c}
\chi _{1}^{0} \\
\chi _{2}^{-} \\
\chi _{3}^{0}%
\end{array}%
\right) \sim \left( \mathbf{1},\mathbf{3},-1/3\right) ,\text{ }\eta =\left(
\begin{array}{c}
\eta _{1}^{0} \\
\eta _{2}^{-} \\
\eta _{3}^{0}%
\end{array}%
\right) \sim \left( \mathbf{1},\mathbf{3},-1/3\right) ,\text{ }\rho =\left(
\begin{array}{c}
\rho _{1}^{+} \\
\rho _{2}^{0} \\
\rho _{3}^{+}%
\end{array}%
\right) \sim \left( \mathbf{1},\mathbf{3},2/3\right).
\end{equation}

These scalar triplets are sufficient to successfully yield masses to all fermions, except neutrinos. The spontaneous symmetry breaking goes as: first  $%
SU(3)_{L}\otimes U(1)_{X}\overset{\left\langle \chi \right\rangle }{%
\rightarrow }SU(2)_{L}\otimes U(1)_{Y},\ $while$\ \ SU(2)_{L}\otimes U(1)_{Y}%
\overset{\left\langle \eta \right\rangle ,\left\langle \rho \right\rangle }{%
\rightarrow }U(1)_{Q}$. In other words, when $\chi $ develops vacuum expectation values the 3-3-1 symmetry is broken reproducing the SM gauge group, which then breaks into quantum electrodynamics after $\eta$ and $\rho $ acquire a vacuum expectation value. For the purpose of this work, it easier to work on the broken phase where, 
\begin{equation}
\left\langle \chi \right\rangle =\frac{1}{\sqrt{2}}\left(
\begin{array}{c}
0 \\
0 \\
v_{\chi }%
\end{array}%
\right) ,\text{ }\left\langle \eta \right\rangle =\frac{1}{\sqrt{2}}\left(
\begin{array}{c}
v_{\eta } \\
0 \\
0%
\end{array}%
\right) ,\text{ }\left\langle \rho \right\rangle =\frac{1}{\sqrt{2}}\left(
\begin{array}{c}
0 \\
v_{\rho } \\
0%
\end{array}%
\right) .
\end{equation}

For simplicity we will adopt $v_{\eta}=v_{\rho}$. This assumption is typically made to simplify the diagonalization of the mass matrices involving the scalars \cite{Mizukoshi:2010ky}. Moreover, we take $v_{\eta}^2+v_{\rho}^2 \sim 246^2 {\rm GeV^2}$, this ought to be enforced to generate the correct masses for the W and Z bosons in the SM, as occurs in models with extra scalars contributing to gauge boson masses \cite{Camargo:2018uzw,Camargo:2019ukv}. \\

We have explained in detail how each mass is generated in the {\it Appendix}, for this reason, in what follows we will just briefly consider each sector separately.

\subsubsection{Charged Leptons Masses}

The masses for the charged leptons are generated via the lagrangian,
\begin{equation}
\mathcal{L}_{Y}^{l}=h_{ab}^{l}\overline{\psi }_{aL}\rho e_{bR} +h.c.\,.
\label{yukawa1}
\end{equation} 

Notice that this term conserves lepton number. One possible term that one could write down is $\epsilon_{ijk}\, \overline{\psi^c}_{L\,i} \psi_{Lj} \rho_k$, which does not conserve lepton flavor. The reason why we did not include this term in Eq.\eqref{yukawa1} is that there is a set of discrete symmetries that will invoke later on to prevent mixing between the SM quarks and the exotic ones. One of them requires $\rho \rightarrow -\rho$, which forbids the term above. However, we need to impose $e_R \rightarrow - e_R$ to engender masses for charged leptons. When the field $\rho_2^0$ develops a vacuum expectation value, $v_\rho$, we find,

\begin{equation}
m_l= h_{ab}^l \frac{v_\rho}{\sqrt{2}}.
\end{equation}

\subsubsection{Quarks Masses}

The quark masses could stem from two sources, one where flavor is conserved (LFC) and other where it is violated (LFV) as follows,
\begin{eqnarray}
\mathcal{L}_{LFC} &=&h_{\alpha a}^{u}\overline{Q}_{\alpha L}\rho ^{\ast
}u_{aR}+h_{\alpha a}^{d}\overline{Q}_{\alpha L}\eta ^{\ast }d_{aR}+h^{U}%
\overline{Q}_{3L}\chi U_{R}  \notag \\
&&+h_{a}^{d}\overline{Q}_{3L}\rho d_{aR}+h_{a}^{u}\overline{Q}_{3L}\eta
u_{aR}+h_{\alpha \beta }^{D}\overline{Q}_{\alpha L}\chi ^{\ast }D_{\beta
R}+h.c.,  \label{eqLFC}
\end{eqnarray}and 

\begin{eqnarray*}
\mathcal{L}_{LFV} &=&s_{a}^{u}\overline{Q}_{3L}\chi u_{aR}+s_{\alpha a}^{d}\overline{Q}%
_{\alpha L}\chi ^{\ast }d_{\beta R}+s^{U}\overline{Q}_{3L}\eta U_{R} \\
&&+s_{\alpha a}^{D}\overline{Q}_{\alpha L}\eta ^{\ast }D_{aR}+s_{\alpha }^{D}%
\overline{Q}_{3L}\rho D_{\alpha R}+s_{\alpha }^{U}\overline{Q}_{\alpha
L}\rho ^{\ast }U_{R}+h.c..
\end{eqnarray*}%

Nevertheless, the LFV lagrangian is problematic because it induces mixing between the SM quarks and the exotic ones. This mixing could alter the properties of the SM quarks. Thus we need to eliminate them. To do so, we invoke a set of $Z_2$ symmetries where some fields are odd under: $u_{aR} \rightarrow -u_{aR}$, $d_{aR}\rightarrow - d_{aR}$, $D_{aR} \rightarrow - D_{aR}$, $\eta \rightarrow -\eta$, $\rho \rightarrow -\rho$. The reaming fields transform trivially. We emphasize that these discrete symmetries do not affect the other sectors of the model that is still capable of reproducing the SM features at low energy scales with no prejudice. \\

In summary, with these discrete symmetries the SM quarks get masses through Eq.\eqref{eqLFC} yielding,

\begin{equation}
m_{u,c} =  \frac{-h^u_{11,22} v_{\rho} }{\sqrt{2}},\quad m_{t} =  \frac{h^u_{33} v_{\eta} }{\sqrt{2}},  
\end{equation}and,

\begin{equation}
m_{d,s} =  \frac{h^d_{11,22} v_{\eta} }{\sqrt{2}},\quad m_{b} =  \frac{h^d_{33} v_{\rho} }{\sqrt{2}}.  
\end{equation}

It is visible that their masses are proportional to $v_{\eta}$ and $v_{\rho}$ which are set at the electroweak scale to avoid flavor changing interactions and return the correct quark masses \cite{Rodriguez:2004mw,Cogollo:2012ek,Giang:2012vs,Cogollo:2013mga,Machado:2013jca,Boucenna:2015zwa,Vien:2015koa,CarcamoHernandez:2018iel}.

Moreover, from Eq.\eqref{eqLFC} we find the exotic quarks masses,
\begin{equation}
m_U =\frac{h^U v_\chi}{\sqrt{2}},\quad m_D =\frac{h^D_{11,22} v_\chi}{\sqrt{2}}  
\end{equation}which are  proportional to the scale of the 3-3-1 symmetry breaking, $v_\chi$, assumed to be sufficiently high to be in agreement with null results from collider searches for exotic quarks \cite{Dobrescu:2016pda,Chala:2017xgc,Serkin:2019jiv}. We will now turn our attention to neutrino masses.

\subsubsection{Neutrino Masses- Type $I+II$ Seesaw}

As far as neutrino masses are concerned, the scalar triplets do not suffice to successfully generate neutrino masses. Using Lie algebra one may notice that $\bar{\psi_L^c} \psi_L \sim (\mathbf{3}^{\ast} + \mathbf{6}, -2/3)$. Thus, one may generate neutrino masses via the scalar triplet $\rho$ and a scalar sextet \cite{Foot:1994ym,Ky:2005yq,Dong:2008sw} with,

\begin{equation}
S=\left(
\begin{tabular}{lll}
$S_{11}^{0}$ & $S_{12}^{-}$ & $S_{13}^{0}$ \\
$S_{12}^{-}$ & $S_{22}^{--}$ & $S_{23}^{-}$ \\
$S_{13}^{0}$ & $S_{23}^{-}$ & $S_{33}^{0}$%
\end{tabular}%
\right) \sim \left( \mathbf{1},\mathbf{6},-2/3\right).
\label{eqsextet}
\end{equation}

However, the discrete symmetries mentioned previously prohibit the Yukawa term proportional to $\rho$ and for this reason, only the scalar sextet contributes leading to,

\begin{equation}
\mathcal{L}_{Y}^{l}=f_{ab}^{\nu }(\overline{\psi }_{aL}%
)_{m}(\psi _{bL}^{c})_{n}S_{mn}+h.c.\, .
\label{yukawanu}
\end{equation}%

Neutrino masses arise after the neutral scalars develop vacuum expectation value as follows, 

\begin{equation}
\left\langle S\right\rangle =\frac{1}{\sqrt{2}}\left(
\begin{tabular}{lll}
$v_{s_{11}}$ & $0$ & $v_{s_{13}}$ \\
$0$ & $0$ & $0$ \\
$v_{s_{13}}$ & $0$ & $\Lambda $%
\end{tabular}%
\right)
\end{equation}%

Notice that $v_{s11},v_{s13}$, and 
$\Lambda$ are the vacuum expectation value of the neutral scalars in the sextet.

With this vacuum structure we generate a mass matrix of the form,

\begin{equation}
\mathcal{L}_{mass}^{\nu }=-\frac{1}{2}\left(
\begin{array}{cc}
\overline{\nu }_{aL} & \overline{\nu }_{aR}^{c}%
\end{array}%
\right) M_{\nu }\left(
\begin{array}{c}
\nu _{bL}^{c} \\
\nu _{bR}%
\end{array}%
\right) +h.c.
\label{nuLagran}
\end{equation}%
where $M_{\nu }$,
\begin{equation}
M_{\nu }=\left(
\begin{array}{cc}
M_{L} & M_{D} \\
M_{D}^{T} & M_{R}%
\end{array}%
\right)
\label{matrixnu}
\end{equation}with,

\begin{eqnarray}
M_{L} &=&\sqrt{2}v_{s_{11}}f^{\nu }, \\
M_{D} &=&\sqrt{2} v_{s_{13}}f^{\nu } , \\
M_{R} &=&\sqrt{2}\Lambda f^{\nu }.
\label{massterms}
\end{eqnarray}%

We remind the reader that $M_{\nu }$ is a $6\times 6$
matrix and $M_{L},M_{D}$ and $M_{R}$ as $3\times 3$ matrices. After the diagonalization procedure we get for the active neutrinos,
\begin{equation}
(m_{\nu})_{ab}=\sqrt{2}\left( v_{s_{11}}-\frac{v_{s_{13}}^{2}}{\Lambda }\right) (f^{\nu
})_{ab},
\label{massnu}
\end{equation}and for the right-handed ones,

\begin{equation}
(M_{\nu_R})_{ab}=\sqrt{2}\Lambda (f^{\nu })_{ab}.
\label{massNR}
\end{equation}\\

It is good timing to highlight a few things: \\

(i) the mass ratio $m_\nu/m_{\nu_R}$ is independent of the Yukawa coupling;\\ 

(ii) a scalar sextet breaks down to a triplet plus doublet plus singlet scalar, i.e. $\mathbf{6}\rightarrow \mathbf{3}+\mathbf{2}+\mathbf{1}$. In order words, the scalar sextet generates the triplet scalar used in the type II seesaw, the doublet scalar that induces Dirac masses, $M_D$, and a singlet scalar that yields right-handed Majorana masses, $M_R$. Therefore, the scalar sextet naturally gives rise to a type I +II seesaw mechanism via spontaneous symmetry breaking; \\

(iii) Setting $v_{s13}^2/\Lambda \sim v_{s11}$ the smallness of the active neutrino masses are justified by taking $v_{s11}$ to be around $1$~eV. \\

There are different possibilities to successfully generate neutrino masses. Since the vacuum expectation values of the fields in the scalar sextet are in principle arbitrary one can play with them and find different Yukawa couplings leading to the same neutrino masses. Be that as it may, one can draw important and insightful conclusions by adopting some simplifications. We will assume throughout that  $v_{s13}^2/\Lambda \sim v_{s11}$ and $v_{s13}=v_{\rho}=v_{\eta}$. Therefore, if we choose $v_{s11}=1$~eV this automatically translates into $\Lambda\sim 10^{13}$~GeV, and $v_{s11}=100$~eV yields $\Lambda\sim 10^{11}$~GeV. Later on, we will present several benchmark scenarios taking $v_{s11}=1$~eV and $v_{s11}=100$~eV to investigate the impact of the neutrino mass ordering on the lepton flavor violating muon decays. With this information, one can now have an estimate of the right-handed neutrino masses since they are proportional to $\Lambda$.\\

For simplicity we will adopt that $v_{s13}^2/\Lambda\sim v_{s11}$ so that $(m_{\nu})_{ab} =\sqrt{2} f^{\nu}_{ab}$. In this way, we have a dominant type II seesaw mechanism with the neutrino masses governed by the vacuum expectation value of $S^0_{11}$. One can think of it as the vacuum expectation value of the triplet under $SU(2)_L$ in the broken phase in the usual type II seesaw study. \\

 Looking at Eq.\eqref{yukawanu} we notice that $S^0_{33},S^0_{11},S^-_{12},S^{--}_{22}$ carry two units of lepton number and for this reason are called bileptons. The lepton flavor (number) is violated after the neutral components $S^0_{11}$ and $S^0_{33}$ acquire a vacuum expectation value. The singly charged scalar in the scalar sextet will not mix with other singly charged scalars that do not carry lepton number. The presence of flavor violation will be explored in this work in the context of muon decays.

\subsection{Gauge Sector}

We turn our attention to the gauge sector. The main point is to show that there are new gauge bosons with masses are proportional to the 3-3-1 scale of symmetry breaking and they are subject to stringent collider bounds. Showing this in a pedagogical manner requires us to start with the covariant derivative of $SU(3)_{L}\otimes
U(1)_{N}$,
\begin{equation}
\mathcal{D}_{\mu }\varphi =\left[ \partial _{\mu }-ig_{L}W_{\mu }^{m}\frac{%
\lambda ^{m}}{2}-ig_{N}N_{\varphi }W_{\mu }^{N}\right]\varphi,
\end{equation}where $g_{L}$ is the coupling constant of $SU_{L}\left( 3\right)$ group, $g_{N}$ is the coupling constant of $U(1)_{N},$ $\lambda ^{m}$ are the Gell-Mann Matrices with $m=1,...,8,$ $W_{\mu }^{m}$ are the gauge
bosons in the adjoint representation of $SU(3)_{L}$, $ W_{\mu}^{N}$ is the gauge field associated to  $U(1)_{N}$ and $N_{\varphi }$ is the
hypercharge associated to $U(1)_{N}$. Writing down the term proportional to $\lambda^m$ one finds, 

\begin{equation}
\frac{g_{L}}{2}W_{\mu }^{m}\lambda ^{m}=\frac{g_{L}}{2}\left(
\begin{array}{ccc}
\left( W_{\mu }^{3}+\frac{1}{\sqrt{3}}W_{\mu }^{8}\right) & \sqrt{2}W_{\mu
}^{+} & \sqrt{2}U_{\mu }^{0} \\
\sqrt{2}W_{\mu }^{-} & \frac{g_{L}}{2}\left( -W_{\mu }^{3}+\frac{1}{\sqrt{3}}%
W_{\mu }^{8}\right) & \sqrt{2}W_{\prime \mu }^{-} \\
\sqrt{2}U_{\mu }^{0\dagger } & \sqrt{2}W_{\prime \mu }^{+} & -\frac{2}{\sqrt{3}}%
W_{\mu }^{8}%
\end{array}%
\right) ,
\label{eqbosons}
\end{equation}where $W_{\mu }^{\pm }=\left( W_{\mu }^{1}\mp iW_{\mu }^{2}\right) /\sqrt{2}%
, $ $U_{\mu }^{0}=\left( W_{\mu }^{4}+iW_{\mu }^{5}\right) /\sqrt{2},U_{\mu
}^{0\dagger }=\left( W_{\mu }^{4}-iW_{\mu }^{5}\right) /\sqrt{2},$ $W_{\mu
}^{\prime \pm }=\left( W_{\mu }^{6}\pm iW_{\mu }^{7}\right) /\sqrt{2}$.\\

Therefore, the {\it 3-3-1 model with right-handed neutrinos} predicts the existence of new charged gauge bosons, $W^{\prime \pm}$, which are subject to intense searches at the LHC \cite{Lindner:2016lpp,Lindner:2016lxq}, and exotic neutral gauge bosons $U^{0}$ and $U^{0\dagger}$ \cite{RamirezBarreto:2008zz}. Moreover, from a combination of the $W_8$, $W_3$ and $W_N$ fields, we extract the SM photon and $Z$ bosons, as well as a massive $Z^{\prime}$ field. It is clear the model add five new gauge bosons to the SM, as a direct result of the extended gauge sector which predicted $N^2-1$ bosons, where $N=3$ in this case. \\

The masses of SM gauge bosons are slightly altered by the presence of the scalar sextet. This change is proportional to the vacuum expectation value $v_{s11}$ which is meant to be small. A similar conclusion is found in the usual type II seesaw mechanism. The bound that rises from the $\rho$ parameter enforces \cite{Camargo:2018uzw} at $3\sigma$,

\begin{equation}
v_{s_1} \leq 2 GeV,    
\end{equation}

The masses of the new gauge bosons are all proportional to the scale of symmetry breaking of the model. For a complete spectrum and equations of the gauge boson masses, we refer to \cite{Mizukoshi:2010ky}. Although, some relations are quite useful to ease our reasoning. One can find that,

\begin{equation}
m_{Z^\prime} \simeq 0.4 v_{\chi},\, m_{W^\prime} \simeq 0.32 v_{\chi}
\label{Zprimemass}
\end{equation}i.e. $m_{Z^\prime} \simeq 1.25\, m_{W^\prime}$. \\

Hence, a lower mass bound on the $Z^\prime$ boson represents a direct lower mass bound on the $W^\prime$, similarly to what occurs in the minimal left-right model \cite{Lindner:2016lpp,Lindner:2016lxq}. One may wonder about the existence of important limits related collider searches for the $U^0$ gauge boson. Although, its mass is identical to the $W^\prime$ which is subject to much more intensive searches. At the end of the day, the most important gauge boson as far as collider searches are concerned will be the $Z^\prime$ field, as we explain further.

\section{Collider Bounds}

\subsection{Searches for $Z^\prime$ bosons }

Collider searches for heavy $Z^\prime$ bosons are quite popular because they typically feature a clear signal. If they couple to fermions and have a narrow width, they give rise to pronounced bumps in the dilepton or dijet invariant mass \cite{Alves:2015pea}. If the couplings to leptons are not very suppressed, the use of dilepton data is more promising because it is subject to a smaller SM background. In this model, the $Z^\prime$ couplings to leptons are not small. Using LHC data at $13$~TeV center-of-mass energy with $3.2$ $fb^{-1}$ of integrated luminosity the authors in \cite{Alves:2016fqe} placed a lower mass bound of $3$~TeV. Later, in \cite{Lindner:2016bgg} this limit was improved using $36$ $fb^{-1}$ and $3$ $ab^{-1}$ of integrated luminosity finding,

\begin{equation}
m_{Z^\prime}  >  4~{\rm TeV}\,\, (\mathcal{L}=36~fb^{-1}) ,\quad m_{Z^\prime}  >  6.4~{\rm TeV} \,\, (\mathcal{L}=3~ab^{-1}).
\end{equation}

We emphasize that this limit of $4$~TeV relies on the dielectron plus dimuon data with invariant mass in the $500-6000$~GeV mass range as recommended, with the cuts in transverse energy and momentum as recommended by ATLAS collaboration in \cite{Aaboud:2016ejt,Aaboud:2017buh}. The projected bound of $6.4$~TeV assumes a similar detector with the same trigger efficiency running at $14$~TeV center-of-mass energy and with $3 ab^{-1}$ of integrated luminosity. This lower mass bound of $4$~TeV is rather robust and important because the $Z^\prime$ is tied to the scale of symmetry breaking of the model. Since $m_{Z^\prime}= 0.4 v_{\chi}$ (see Eq.\eqref{Zprimemass}), we automatically find $v_{\chi}> 10$~TeV.  Furthermore, using the mass relations between the $W^\prime$ and $Z^\prime$ bosons one finds a lower mass bound on the $W^\prime$ mass which is much restrictive than other \cite{RamirezBarreto:2008wq,RamirezBarreto:2011av} on its mass. Anyway, the $Z^\prime$ bound aforementioned is the most relevant to our discussion. We highlight that this limit has nothing to do with lepton flavor violation, it is rather simply based on searches for heavy dilepton resonances. Therefore, they are not relevant to our discussion because the mass of the scalars that enter in the lepton flavor violation discussion are not much sensitive to the energy scale of 3-3-1 symmetry breaking. Nevertheless, we emphasize that our entire discussion of lepton flavor violation is fully consistent with these bounds on the $Z^\prime$ mass. Now we will address collider searches for lepton flavor violation in what follows.  

\subsection{Searches for a Doubly Charged Scalar at the LHC}

The presence of a doubly charged scalar in the spectrum is typical signature of a type II seesaw mechanism \cite{Kanemura:2009mk,Majee:2010ar,Han:2015hba,Chen:2016xju,Ghosh:2017pxl,Dev:2018sel,Li:2018abw,Li:2018jns,Ouazghour:2018mld}. Typically, this type II seesaw is realized by the addition of a scalar triplet. This popular extension triggered several phenomenological analyses. As we mentioned previously, the scalar triplet that arises after the spontaneous symmetry breaking is key to the type II seesaw mechanism in our model. We emphasize, however, that in our model we have a type I + II seesaw because we do also have Dirac neutrino masses. After spontaneous symmetry breaking, we can single out the Yukawa term involving the scalar triplet and SM particles which reads \cite{Camargo:2018uzw}, 

\begin{equation}
\mathcal{L} = f_{ab}^\nu \bar{L}^c_{a} \Delta^{\ast} L_b   
\end{equation}where $L$ is the SM lepton doublet with,

\begin{center}
\begin{equation}
   \Delta=\left(\begin{array}{cc}
        S_{}^0 & S_{}^-  \\
       S_{}^-  &  S^{--}\\
    \end{array}\right),
\end{equation}
\end{center}where $S^0\equiv S^0_{11}$, $S^-\equiv  S^-_{12}$, $S^{--}\equiv  S_{22}^{--}$, as defined in Eq.\eqref{eqsextet}.\\

The doubly charged scalar decay width into charged leptons is found to be \cite{CMS:2017pet},
\begin{equation}
\Gamma(S^{\pm\pm} \rightarrow l^{\pm}_a l^{\pm}_b )= 1/(2 \pi x)  |f^\nu_{ab}|^2 m_{S^{\pm\pm}}^2,
\end{equation}where $x= 1$ for $i \neq j$ and $x=2$ for $i=j$. Thus is clear that the branching ratio into charged leptons can change depending on the Yukawa couplings. Different choices lead to different branching ratios and consequently different lower mass bounds. The dominant decay mode determines cuts, detector efficiency, and backgrounds which the signal is subject to, and consequently yielding different lower mass bounds. Dielectron and dimuon channels offer a cleaner environment and thus yield stronger bounds. If lepton flavor violation is assumed the SM background is suppressed, which again strengthens the limits. This is reasoning behind the limits derived using LHC data.\\

We used the code {\it fastlim} described in \cite{Papucci:2014rja}, and adopted a parton distribution function at next-to-next leading \cite{Kramer:2012bx,Borschensky:2014cia} in order to project the LHC sensitivity for a high-luminosity (HL) setup, following the recommendations presented in \cite{CidVidal:2018eel}. We stress that the HL-LHC limit refers to a detector similar to LHC running at $14$~TeV with $3~ab^{-1}$ of integrated-luminosity. The High-Energy LHC configurations represents a $27$~TeV colliding beam with $15~ab^{-1}$ of data. We highlight that these bounds are based on the simulated signal $q\bar{q} \rightarrow Z,\gamma \rightarrow \phi^{++}\phi^{--}$ as outlined in \cite{CMS:2017pet}, which features bounds stronger than previous previous studies \cite{Chatrchyan:2012ya,Aaboud:2017qph,Aaboud:2018qcu}.  In summary we derived,

\begin{eqnarray}
m_{S^{\pm\pm}} & >& 943\,{\rm GeV} \,\, (\text{LHC}),\nonumber\\ m_{S^{\pm\pm}} & > &  2.5\, {\rm TeV}\,\,  (\text{HL-LHC})\nonumber\\
m_{S^{\pm\pm}} & > &  4.9\, {\rm TeV}\,\, (\text{HE-LHC})
\label{boundLHC}
\end{eqnarray}

The latest LHC search for doubly charged scalars was with $12.9~fb^{-1}$ of data, but notice that with $36~fb^{-1}$ we expect LHC to already rule out doubly charged scalars with masses around $940$~GeV, assuming a normal mass ordering for the active neutrinos. If we had considered an inverted mass ordering a weaker bound would have been found, lying around $900$~GeV. This difference in the lower mass bound from LHC for normal and inverted ordering will not cause a meaningful impact on our conclusions and with this understanding in mind, we will simply quote those in Eq.\eqref{boundLHC}. There are other important limits on this scenario \cite{Dev:2017ouk,Agrawal:2018pci,Dev:2018sel,Du:2018eaw,Dev:2019hev} but the ones we quote are the most relevant.  For a more complete discussion of the collider bounds on the type II seesaw we refer to \cite{Primulando:2019evb}.\\

Moreover, it is exciting to see that HE-LHC can potentially probe doubly charged scalar with masses up to $\sim 5$~TeV. These limits are quite important and serve as an orthogonal test to the type I + type II seesaw scenario we are investigating because that restricts the region in which the seesaw mechanism is viable. Having in mind that the doubly charged scalar is key to the lepton flavor violation observables we are about to discuss, such collider bounds stand as a complementary and important cross-check to lepton flavor violation signatures. 

\section{Lepton Flavor Violation}

Lepton flavor violation is one of the most interesting probes of physics beyond the SM. The main lepton flavor violation signatures of the seesaw mechanism stem from muon decay namely, $\mu \rightarrow e\gamma$ and $\mu \rightarrow 3e$. There are other sources of lepton flavor violation such as $\mu-e$ conversion but they are subdominant \cite{Lychkovskiy:2010ue,Dinh:2012bp,Abada:2014kba,He:2014efa}. Other lepton flavor violating decays involving the $\tau$ lepton are less promising, unless one invokes a mechanism to significantly suppress $\mu \rightarrow e\gamma$ \cite{Matsuzaki:2008ik,Akeroyd:2009nu,Celis:2013xja,Hue:2013uw,Dinh:2013vya,Omura:2015xcg,Zhou:2016ynv,Dib:2018rpy,Hernandez-Tome:2018fbq,Calcuttawala:2018wgo}. Anyway, going back to the relevant muon decays, $\MEG$ and $\mu \rightarrow 3e$, one can check that the current bounds read
 $BR(\MEG) < 4.2 \times 10^{-13}, BR(\mu \rightarrow 3e) < 10^{-12}$, and future experiments aim $BR(\MEG) < 4 \times 10^{-14}, BR(\mu \rightarrow 3e) < 10^{-16}$. Therefore, we expect an important experimental improvement in the near future. Eventually, we will superimpose these limits with the model's contribution. That said, having in mind that we have a dominant type II seesaw setup, the first contribution to $\MEG$ in our models stems from 1-loop processes involving the doubly and singly charged scalars which lead to, 
\begin{equation}
BR(\mu \rightarrow e\gamma )\simeq \frac{\alpha _{EM}\left\vert
(f_{ab}^{\ast }\text{ }f_{ab})_{e\mu }\right\vert ^{2}}{192\pi G_{F}^{2}\text{
}}\left( \frac{1}{m_{S^{\pm\pm}}^{2}}+\frac{8}{m_{S ^{\pm}}^{2}}\right) ^{2},
\label{eqBmue}
\end{equation}%
where $\alpha _{EM}$ is the fine-structure constant, $G_{F}$ the Fermi
constant, $m_{S^{\pm\pm}}$ the mass of doubly charged scalar, $m_{S^{\pm}}$ the mass of singly charged scalar. In what follows will assume that the doubly charged and singly charged scalar have the same mass. This assumption will allow us to connect $\MEG$ directly to $\mu\rightarrow$ 3e and LHC limits on the doubly charged scalar mass.  

There is an additional source of lepton flavor violation that rises from charged current which reads,
\begin{equation}
\mathcal{L} = \frac{g}{2\sqrt{2}} \bar{l} \gamma^\mu (1-\gamma_5) (\nu_R)^c W^{\prime-},
\label{eq.LFV2}
\end{equation}where $W^{\prime}$ , is a charged gauge boson defined in Eq.\eqref{eqbosons} and has a mass equal to $0.3 v_{\chi}$, $v_{\chi}$ being the energy scale of 3-3-1 symmetry breaking. This interaction results in the following branching ratio,
\begin{equation}
{\rm BR} (\mu \rightarrow e\gamma)= \frac{3 (4\pi)^3 \alpha_{em}}{4 G_F^2}\left( |A^M_{e\mu}|^2 + |A^{E}_{e\mu}|^2\right)
\label{MEGgeneral}
\end{equation}where,
\begin{eqnarray}
\label{Eq:similartoseesawmutoe}
A^M_{e\mu} =& \frac{-1}{(4\pi)^2}\sum_f \left( {g_{v\,1}^{fe}}^*g_{v\,1}^{f\mu} I_{f,\, 3}^{+\,+}+{g_{a\,1}^{fe}}^*g_{a\,1}^{f\mu} I_{f,\, 3}^{+\,-}\right),\\
A^E_{e\mu} =& \frac{i}{(4\pi)^2}\sum_f \left({g_{a\,1}^{fe}}^*g_{v\,1}^{f\mu} I_{f,\, 3}^{-\,+} + {g_{v\,1}^{fe}}^*g_{a\,1}^{f\mu} I_{f,\, 3}^{-\,-}\right),
\end{eqnarray}with $g_v$, $g_a$ being the couplings constants that encompass the constants in Eq.\eqref{eq.LFV2} and the neutrino mixing matrices, and  $I_{f,\,3}^{\pm\,\pm}$ given by,
\begin{align} \label{eq:I3_def}
  I_{f,\,3}^{(\pm)_1\,(\pm)_2} &\equiv I_{f,\,3}\left[m_i,(\pm)_1 m_j,(\pm)_2 m_{N_f},m_W\right] \nonumber\\
  =& \int \mathrm{d}^3\mathbf{X} \Bigg[\left[-xz\, m_i^2 - xy\, m_j^2 +(1-x) m_W^2 + x\, m_{N_f}^2\right]^{-1} \nonumber\times\\
  & \times\bigg\lbrace -(\pm)_2 3 (1-x) \frac{m_{N_f}}{m_i} + (y+2z(1-x)) +(\pm)_1 \frac{m_j}{m_i}(z+2y(1-x)) \nonumber \\
  & +  \frac{m_i^2}{m_W^2}\bigg[ x\left( (\pm)_1 (1-y)\frac{m_j}{m_i}-z\right)\left(z +(\pm)_1 y\,\frac{m_j}{m_i} +(\pm)_2 \frac{m_{N_f}}{m_i}\right)\left( (\pm)_1 y\,\frac{m_j}{m_i}-(1-z)\right) \nonumber \\
  & + xy\left(1 -(\pm)_2  \frac{m_{N_f}}{m_i}\right)\left(\frac{m_j^2}{m_i^2}\,(1-y) -z\right) \nonumber\\
  &  + xz\left(+(\pm)_1 \frac{m_j}{m_i} -(\pm)_2  \frac{m_{N_f}}{m_i}\right)\left((1-z)-y\frac{m_j^2}{m_i^2}\right) \bigg] \bigg\rbrace \nonumber\\
  & + m_W^{-2} \left[ x (1-z) (\pm)_1 x(1-y) \frac{m_j}{m_i} -(\pm)_2  x \frac{m_{N_f}}{m_i} + y\left(1 -(\pm)_2  \frac{m_{N_f}}{m_i}\right)\right. + \nonumber\\
  & + \left.z \left((\pm)_1  \frac{m_j}{m_i} -(\pm)_2  \frac{m_{N_f}}{m_i}\right)\right] \nonumber\\
  & - m_W^{-2} \bigg[ (1 - 3 x) \left((\pm)_2\frac{m_{N_f}}{m_i} - (1-z) -(\pm)_1 (1 - y) \frac{m_j}{m_i}\right) - x z -(\pm)_1 x y \frac{m_j}{m_i}  \nonumber\\ 	
   &+ \left((\pm)_2\frac{m_{N_f}}{m_i} - 1\right)(1 - 3 y) \left((\pm)_2\frac{m_{N_f}}{m_i} -(\pm)_1 \frac{m_j}{m_i}\right) (1 + 3 z) \bigg] \times \nonumber\\
   & \times\log \left(\frac{m_W^2}{-xz\, m_i^2 - xy\, m_j^2 +(1-x) m_W^2 + x\, m_{N_f}^2}\right) \Bigg] 
\end{align}

As one can see, it is not obvious the computation of the $\MEG$ decay in our model. In the regime which the right-handed neutrinos have identical masses to the $W^{\prime}$, the prediction for $\MEG$ is greatly simplified yielding,

\begin{equation}
{\rm BR} (\MEG) = 1.6 \left(\frac{{\rm 1~TeV}}{m_{W^{\prime}} }  \right)^4  |g^{\nu_R e \ast}  g^{\nu_R \mu}|^2
\end{equation}where $g^{\nu_R e}=g/(2\sqrt{2}) U^{\nu_R e}$ and $g^{\nu_R \mu }=g/(2\sqrt{2}) U^{\nu_R \mu}$. \\

Keeping right-handed neutrino masses similar to the $W^\prime$ mass and around the weak scale one can have a visible $\MEG$ decay, but if right-handed neutrino masses are much larger than the TeV scale, one can plugging the numbers in Eq.\eqref{MEGgeneral} using  Eq.\eqref{Eq:similartoseesawmutoe}-\eqref{eq:I3_def} to show that ${\rm BR} (\MEG) < 10^{-15}$, which is beyond reach current and projected experiments. On the other hand, even in this scenario where right-handed neutrinos are very heavy, the ${\rm BR} (\MEG)$ can be large due to the presence of a type II seesaw mechanism which features doubly charged and singly charged scalars contributions. \\

Bearing in mind that the right-handed neutrinos in our model will be heavy, and one can neglect the $W^\prime$ contribution to the $\MEG$ decay and focus on the seesaw component. That said, another important observable is the $\mu\rightarrow 3e$ decay \cite{Lindner:2016bgg}. Since only the doubly charged Higgs contributes to this decay the calculation is simpler and leads to,

\begin{equation}
{\rm BR} (\mu \rightarrow 3e) =\frac{| f^{\nu\dagger}_{ee} f^{\nu}_{\mu e}|^2 }{G_F^2 m_{S^{\pm\pm}}^4}.  
\label{eqBR3e}
\end{equation}

On one hand we can see that the Yukawa couplings $f^{\nu}_{ab}$ dictate the lepton flavor violation observables, but on the other hand these Yukawa couplings enter in the neutrino mass matrix. Therefore, neutrino masses and lepton flavor violation observables are correlated.\\

Going back to Eq.\eqref{massnu}, if $v_{s13}^2/\Lambda \sim v_{s11}$ as we will assume throughout, we have a dominant type II seesaw setup with  the active neutrino masses set by vacuum expectation value, $v_{s11}$, i,e. $(m_\nu)_{ab} \simeq \sqrt{2} f^\nu_{ab} v_{s11}$. The picture is not so simple because neutrinos oscillate and we need to reproduce the oscillation pattern. Therefore, these Yukawa couplings are found to be \cite{Dinh:2012bp},

\begin{equation}
f^{\nu}_{ab}=\frac{1}{\sqrt{2}v_{s11}}(U^{\ast} diag(m_{\nu1},m_{\nu2},m_{\nu3}) U^{\dagger})_{ab}
\label{fabeq}
\end{equation}where U is the Pontecorvo, Maki, Nakagawa, Sakata (PMNS) neutrino mixing matrix of dimension $3\times3$, parametrized as follows \cite{Tanabashi:2018oca},

\begin{equation}
\left(\begin{array}{ccc}
  1 & 0   & 0 \\
  0  & \cos\theta_{23} & \sin\theta_{23} \\
   0 & -\sin\theta_{23}  & \cos\theta_{23}
\end{array}  \right) \left(\begin{array}{ccc}
  \cos\theta_{13} & 0   & \sin\theta_{13} e^{-i\delta} \\
  0  & 1 & 0 \\
   -\sin\theta_{13} e^{i\delta} & 0 & \cos\theta_{13}
\end{array}  \right) 
\left(\begin{array}{ccc}
  \cos\theta_{12} & \sin\theta_{12}   & 0 \\
 -\sin\theta_{12}  & \cos\theta_{12} & 0 \\
   0 & 0  & 1
\end{array}  \right) 
\end{equation}with the mixing parameters as shown in {\it Table \ref{tabnumixing}}. \cite{Capozzi:2017ipn},

\begin{table}[]
\begin{equation}
 \begin{array}{|c|c|c|}
 \hline
  {\rm Parameter} & {\rm Best-fit} & {\rm Hierarchy} \\
  \hline
   \Delta m_{21}^2   & 7.37\times 10^{-5}~eV^2  & {\rm Any} \\
   \Delta m_{31}^2 & 2.56\times 10^{-3}~eV^2 & {\rm Normal}\\
   \Delta m_{23}^2 & 2.56\times 10^{-3}~eV^2 & {\rm Inverted}\\
   \sin^2\theta_{12} & 0.297 & {\rm Any} \\
   \sin^2\theta_{23} & 0.425& {\rm Normal} \\
    \sin^2\theta_{23} & 0.589& {\rm Inverted} \\
     \sin^2\theta_{13} & 0.0215& {\rm Normal} \\
      \sin^2\theta_{13} & 0.0216& {\rm Inverted} \\
    \hline
 \end{array}   
\end{equation}
\label{tabnumixing}
\caption{Table with the best-fit parameters that enter in the neutrino mass mixing according to \cite{Patrignani:2016xqp}.}
\end{table}

In summary, one needs to incorporate neutrino oscillations to have more solid predictions for the lepton flavor violation observables. We will explore this fact by investigating several benchmark points which encompass normal and inverted mass ordering and different absolute neutrino masses. We will show that the neutrino mass spectrum is rather relevant to the overall lepton flavor violation signatures, a fact that has not been explored in detail in the context of 3-3-1 models. With this input from neutrino oscillations, our reasoning goes as follows:\\

\begin{itemize}
    \item We choose a neutrino mass ordering;
    
    \item Then we pick a neutrino mass $m_{\nu1}$, which then basically fixes $m_{\nu2}$ and $m_{\nu3}$, for a given $vev$, $v_{s11}=1-100$~eV. From this, we find the  Yukawa couplings that reproduce this spectrum taking into account the oscillation patterns using  Eq.\eqref{fabeq};
    \item With these Yukawa couplings we use Eq.\eqref{eqBmue}-\eqref{eqBR3e} to compute the lepton flavor violating muon decays.

\end{itemize}

\begin{figure}[!t]
    \includegraphics[scale=0.65]{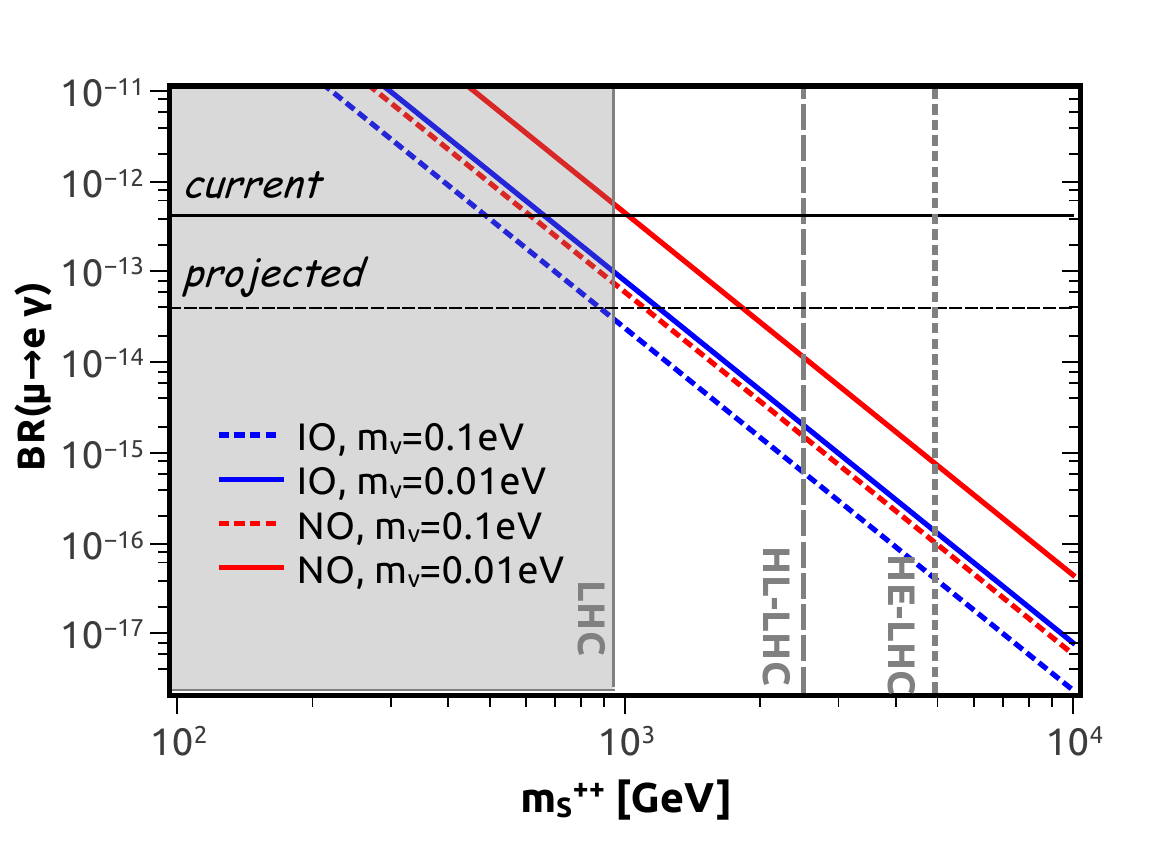}
    \includegraphics[scale=0.65]{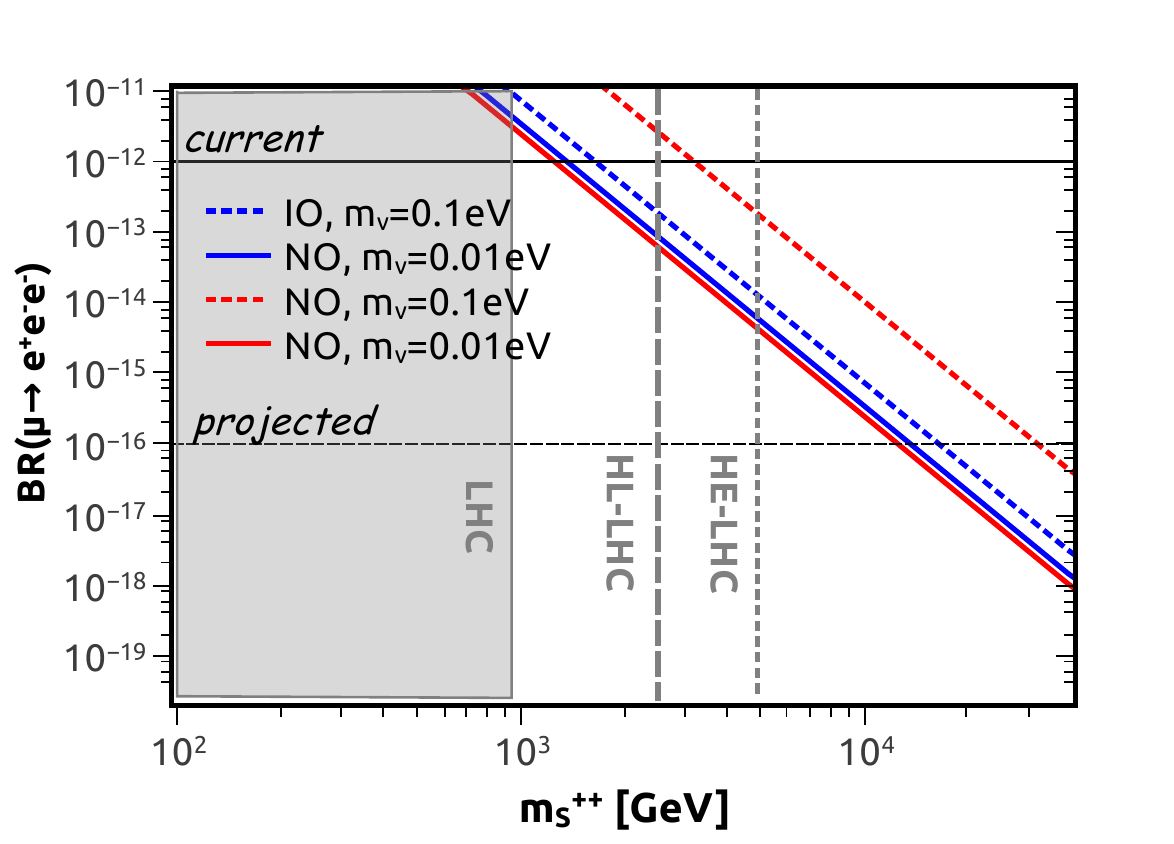}
    \caption{Figures showing the predictions for $\MEG$ ({\it left-panel}) and $\mu\rightarrow 3e$ ({\it right-panel}) decays in our model overlaid with the existing and projected bounds from collider and lepton flavor violation searches.  We used the labels IO (NO) for inverted (normal) mass ordering for the neutrinos. LHC limit refers to the current LHC bound with $36~fb^{-1}$ of data. We also show the HL-LHC, and HE-LHC sensitivities which represent the LHC running with the $14~TeV-3~ab^{-1}$ and $27~TeV-15~ ab^{-1}$ configurations. It is clear from the figures that the neutrino mass ordering a great impact on the theoretical predictions for lepton flavor violation. See text for a detailed discussion.}
    \label{finalfig}
\end{figure}

In {\it Table} \ref{tab:my_label3} we summarize our findings using this logic. The table will allow the reader to easily follow our reasoning as we discuss the results in figure \ref{finalfig} where we display four of these benchmark points for $m_{\nu1}=0.1$~eV and $m_{\nu1}=0.01$~eV including normal and inverted mass ordering. The blue (red) lines are the model predictions for the muon decays for inverted  (normal) neutrino mass hierarchies. The difference between the solid and dashed lines is the absolute mass for the neutrino flavor $\nu_1$. For instance, the red solid line in the {\it left-panel} of figure \ref{finalfig} accounts for the $BR(\MEG)$ for $m_{\nu1}=0.01$~eV within a normal mass ordering. The gray region represents the region currently excluded by LHC based on the search for doubly charged scalars within a type II seesaw framework with normal mass ordering. The dashed and dotted vertical gray lines are the HL-LHC and HE-LHC projected exclusion limits. The black horizontal lines are the current and projected bounds on the $\MEG$ (left-panel) and $\mu\rightarrow 3e$ (right-panel) decays.\\

One can notice that the neutrino mass ordering has a great impact on the lepton flavor violating muon decays. Looking at the first benchmark scenario with $m_{\nu1}=0.1$~eV and $v_{s11}=1$~eV, which assumes an inverted mass ordering (IO) we get BR$(\MEG)= 0.024/m_{S^{\pm\pm}}^4$,  BR$(\mu \rightarrow 3e)= 7.5/m_{S^{\pm\pm}}^4$. Having in mind that the current experimental limits, we conclude that we can probe doubly charged scalars with masses of $600$~GeV and $1.3$~TeV using from $\MEG$ and $\mu\rightarrow 3e$ decays. It is exciting to see that in this setup $\mu \rightarrow 3e$ provides stronger bounds than the LHC. \\

\begin{table}[t]
    \centering
    \begin{tabular}{|c|c|c|c|}
    \hline
      {\bf Benchmark}  & {\bf Hierarchy}  & BR ($\MEG$) & ${\rm BR} (\mu \rightarrow 3e)$\\
       \hline
       $m_{\nu1}=0.1$~eV,~$v_{s11}=1$~eV &  IO  & $0.024/m_{S^{\pm\pm}}^4$& $7.5/m_{S^{\pm\pm}}^4$\\  
       \hline
       $m_{\nu1}=0.1$~eV,~$v_{s11}=100$~eV & IO & $2.4\times 10^{-10}/m_{S^{\pm\pm}}^4$& $7.5\times 10^{-8}/m_{S^{\pm\pm}}^4$\\  
       \hline
       $m_{\nu1}=0.01$~eV,~$v_{s11}=1$~eV & IO  & $0.08/m_{S^{\pm\pm}}^4$& $3.5/m_{S^{\pm\pm}}^4$\\
       \hline
       $m_{\nu1}=0.01$~eV,~$v_{s11}=100$~eV & IO & $8\times10^{-10}/m_{S^{\pm\pm}}^4$& $3.5\times10^{-8}/m_{S^{\pm\pm}}^4$\\  
       \hline
       $m_{\nu1}=0.1$~eV,~$v_{s11}=1$~eV & {\it NO} & $0.44/m_{S^{\pm\pm}}^4$& $106/m_{S^{\pm\pm}}^4$\\  
       \hline
       $m_{\nu1}=0.1$~eV,~$v_{s11}=100$~eV & {\it NO} & $4.4\times10^{-9}/m_{S^{\pm\pm}}^4$ & $1.1\times10^{-6}/m_{S^{\pm\pm}}^4$\\  
       \hline
       $m_{\nu1}=0.01$~eV,~$v_{s11}=1$~eV & {\it NO}  & $0.06/m_{S^{\pm\pm}}^4$& $2.5/m_{S^{\pm\pm}}^4$\\  
       \hline
       $m_{\nu1}=0.01$~eV,~$v_{s11}=100$~eV & {\it NO}  & $6.2\times10^{-10}/m_{S^{\pm\pm}}^4$& $2.5\times10^{-8}/m_{S^{\pm\pm}}^4$\\  
       \hline
    \end{tabular}
    \caption{Table with the theoretical predictions for the $\MEG$ and $\mu\rightarrow 3e$ decay for the type $I+II$ seesaw setup encompassing normal (NO) and inverted (IO) mass ordering. We highlight that the mass of the doubly charged scalar is in $GeV$ units. The neutrino masses in the IO scenario are given by $m_{\nu2}^2=m_{\nu1}^2-\Delta m_{21}^2$, $m_{\nu3}^2=m_{\nu2}^2-\Delta m_{23}^2$, whereas in NO case are  $m_{\nu2}^2=m_{\nu1}^2-\Delta m_{21}^2$, $m_{\nu3}^2=m_{\nu1}^2+\Delta m_{31}^2$.} 
    \label{tab:my_label3}
\end{table}

However, the larger the vacuum $v_{s11}$ the smaller the Yukawa couplings needed to reproduce the same neutrino masses. Hence, when we set $v_{s11}=100$~eV, the predictions change drastically to  BR$(\MEG)= 2.4\times 10^{-10}/m_{S^{\pm\pm}}^4$,  BR$(\mu \rightarrow 3e)= 7.5\times 10^{-10}/m_{S^{\pm\pm}}^4$. For these scenarios where the vacuum of is much larger than $1$~eV, LHC constitute the best probe. For concreteness, in this second case described above, taking $m_{S^{\pm\pm}}= 1000$~GeV, would lead to muon decays much smaller than current and projected sensitivity \cite{Lindner:2016bgg}, making HL-LHC and HE-LHC the best laboratories, since HE-LHC will probe masses of about $\sim 5$~TeV, for instance. All these conclusions are quite visible in figure~\ref{finalfig}.  \\

Considering the normal mass ordering (NO) we conclude that the qualitative statements do not change. Taking $m_{\nu1}=0.1$~eV, $v_{s11}=1$~eV, quantitatively we notice that the while the inverted hierarchy gives BR$(\MEG) =0.024/m_{S^{\pm\pm}}$ we find BR$(\MEG) =0.44/m_{S^{\pm\pm}}$, which is a factor of 20 larger. A larger much decay into $3e$ is also found for the NO compared to the IO (see figure~\ref{finalfig}). An orthogonal way to look at this is by noticing that the region between the current and projected limits delimit a signal region of lepton flavor violation. Looking at both panels of figure~\ref{finalfig} we conclude that doubly charged scalars with masses around $1-2$~TeV might be spotted at the $\MEG$ decay, whereas the $\mu\rightarrow 3e$ will be able to detect such scalars with masses of up to $\sim 30$~TeV. The experimental progress on the search for the $\mu\rightarrow 3e$ decay is remarkable and it will surpass even the HE-LHC regardless of the mass ordering for benchmark scenarios where $v_{s11}\sim 1$~eV.\\

In the {\it left-panel} of figure~\ref{finalfig} one can clearly see the impact of changing the value of the neutrino masses in the $\MEG$ decay. This can be checked by comparing the ratio between the dashed blue and red lines with the solid blue and red lines. This is not true for the $\mu\rightarrow 3e$ decay though (see {\it right-panel} of figure~\ref{finalfig}) . Comparing the NO and IO predictions for $m_{\nu1}=0.01$~eV and $v_{s11}=1$~eV, we find no much difference, this is because the Yukawa couplings relevant for this observable are similar regardless of the neutrino mass ordering. This would continue to be true if we had taken even smaller values for $m_{\nu1}$ because when $m_{\nu1}$ is sufficiently small, the measured mass differences of the neutrino flavors govern by the neutrino mixings and thus the $\mu\rightarrow 3e$ decay. When we take $m_{\nu1}=0.1$~ eV, then the difference in the predictions for NO and IO is noticeable. One can easily use {\it Table} \ref{tab:my_label3} and figure~\ref{finalfig} to validate our conclusions .\\

We can conclude that regardless of the mass ordering and absolute value of the active neutrino masses that HE-LHC will solidly probe the type II seesaw model with respective to its contributions to the $\MEG$ decay. Concerning the $\mu\rightarrow 3e$ decay the situation changes due to the fantastic experimental sensitivity aimed in the near future. The $\mu\rightarrow 3e$ decay will be able to probe this model for doubly charged scalar masses up to 30 TeV, which is way beyond HE-LHC reach, no matter the neutrino mass ordering. If the absolute neutrino masses are much smaller than $0.01$~eV, then $\mu\rightarrow 3e$ decay becomes smaller making HE-LHC still the best probe. 

Our conclusions explicitly show the importance of searching for signs of lepton flavor violation in collider and muon decays. The conclusion about which probe yields stronger bounds depends strongly on the mass ordering adopted, the absolute neutrino masses and which much decay one considers.  In the $1-5$ TeV mass region of the doubly charged scalar lepton flavor violation experiments and colliders offer orthogonal and complementary probes. Thus if a signal is observed in one of the two new physics searches, the other will be able to assess whether is stems from a seesaw framework. \\

In summary, within the {\it 3-3-1 model with right-handed neutrinos} a type I+II seesaw naturally emerges. In the scenario where we have a dominant type II seesaw, the model offers a clear prediction for the $\MEG$ and $\mu\rightarrow 3e $ decays. One may wonder how one could discriminate our model from other type II seesaw proposal and a plausible answer would go as follows: setting aside the type II seesaw, our model predicts the existence of $W^\prime$ and $Z^\prime$ gauge bosons, as well as heavy exotic quarks. The detection of multiple signals consistent with all these particles could serve a discriminator and favor our models over others. 

\section{Conclusions}

We have discussed a model which promotes $SU(2)_L\times U(1)_Y$ to $SU(3)_L\times U(1)_N$. In this extended gauge sector, all fermions get masses via a spontaneous symmetry breaking mechanism that encompasses three scalar triplets but neutrinos. A scalar sextet is added to incorporate neutrino masses. After spontaneous symmetry breaking this scalar sextet breaks down to a doublet and scalar triplet, which play a role in the type I and type II seesaw mechanism. We focus on a scenario of type II seesaw dominance where the relevant lepton flavor violation observables namely, $\MEG$ and $\mu \rightarrow 3e$, are directly tied to neutrino mass ordering and collider bounds on the doubly charged scalar. \\
 
We have explicitly shown how the absolute mass scale and neutrino mass ordering change the model predictions for lepton flavor violation within the type II seesaw framework. Combining the LHC, HL-LHC, HE-LHC sensitivity to doubly charged scalars and the experimental sensitivity to these rare muon decays we concluded that for doubly charged scalar with masses around $1-5$~TeV, these probes are rather complementary.  Moreover, regardless of the mass ordering HE-LHC is expected to solidly test any possible signal seen in these muon decays. \\
 
One may wonder how one could discriminate our model from other types II seesaw proposals and a plausible answer would rely on the existence of $W^\prime$ and $Z^\prime$ gauge bosons, as well as heavy exotic quarks, all predicted in our model. The detection of multiple signals consistent with all these particles could serve a discriminator and favor our models over others. 

\section{Acknowledgments}
The authors thank Diego Cogollo for discussions. MMF acknowledges financial support from CNPq grant 308933/2015-0, and from FAPEMA agency from the grants FAPEMA/UNIVERSAL/00880/15 and FAPEMA/PRONEX 01452-14. 
SK acknowledges support from Fondecyt (Chile) grant  No. 1190845  and CONICYT (Chile)  PIA/Basal FB0821. TM thanks CAPES for the fellowship. FSQ acknowledges support from CNPq grant 303817/2018-6, UFRN, MEC and ICTPSAIFR FAPESP grant 2016/01343-7.
\appendix 

\section{Appendix}

In this appendix will describe in more detail and pedagogical manner some sections of the model which are relevant to our reasoning.

\subsection{Lepton Masses}

\indent
The leptons masses could in principle be generated via the Yukawa lagrangian,%
\begin{equation}
\mathcal{L}_{Y}^{l}=h_{ab}^{l}\overline{\psi }_{aL}\rho e_{bR}+h_{ab}^{\nu }%
\overline{\psi }_{aL}^{c}\psi _{bL}\rho +f_{ab}^{\nu }(\overline{\psi }_{aL}%
\text{ })_{m}(\psi _{bL}^{c}\text{ })_{n}S_{mn}+h.c.,
\end{equation}%
where $h_{ab}^{\nu }$ is an antisymmetric constant coupling matrix and $%
f_{ab}^{\nu }$ is a symmetric constant coupling matrix. The first and third terms conserve the lepton flavor, while the second term violate it. Some the scalars inside the scalar sextet S do have lepton number, for this reason the third term conserves lepton number as well. 

The second term is not problematic but will be removed because a set of discrete that will be invoked. This set of $Z_2$ symmetries are needed to avoid mixing between the SM quarks and the exotic ones. One of them requires $\rho \rightarrow −\rho$, which forbids the second term above. Although, we also need to impose $e_R \rightarrow -e_R$ to generate masses for charged leptons. 

In particular, the charged lepton masses arise from, 
\begin{eqnarray}
L_{mass}^{l} &=&-h_{ab}^{l}\overline{\psi }_{aL}\rho
e_{bR}+h.c.=-h_{ab}^{l}\left(
\begin{array}{ccc}
\overline{\nu }_{aL} & \overline{e}_{aL} & \overline{\nu }_{aR}^{c}%
\end{array}%
\right) \frac{1}{\sqrt{2}}\left(
\begin{array}{c}
0 \\
v_{\rho } \\
0%
\end{array}%
\right) e_{bR}+h.c.  \notag \\
&=&-h_{ab}^{l}\frac{v_{\rho }}{\sqrt{2}}\overline{e}_{aL}e_{bR}+h.c.
\end{eqnarray}%

In this way, in the flavor space the Dirac mass matrix for the charged leptons, at tree level, is given by%
\begin{equation}
m_{l}=h_{ab}^{l}\frac{v_{\rho }}{\sqrt{2}},
\end{equation}%
where $v_{\rho }$ is the vacuum expectation value of the neutral scalar $\rho _{0}$ and $h_{ab}^{l}$ is the coupling constant matrix.

The neutrino masses come from the first second term only when the neutral components of scalar sextet
acquire a vacuum expectation value as follows,
\begin{eqnarray*}
L_{mass}^{\nu S} &=&-f_{ab}^{\nu }\left(
\begin{array}{ccc}
\overline{\nu }_{aL} & \overline{e}_{aL} & \overline{\nu }_{aR}^{c}%
\end{array}%
\right) \frac{1}{\sqrt{2}}\left(
\begin{array}{ccc}
v_{s_{11}} & 0 & v_{s_{13}} \\
0 & 0 & 0 \\
v_{s_{13}} & 0 & \Lambda%
\end{array}%
\right) \left(
\begin{array}{c}
\left( \nu _{bL}\right) ^{c} \\
\left( e_{bL}\right) ^{c} \\
\left[ \left( \nu _{b}^{c}\right) _{L}\right] ^{c}%
\end{array}%
\right) +h.c. \\
&=&-f_{ab}^{\nu }\frac{v_{s_{11}}}{\sqrt{2}}\overline{\nu }_{aL}\left( \nu
_{bL}\right) ^{c}-f_{ab}^{\nu }\frac{v_{s_{13}}}{\sqrt{2}}\overline{\nu }_{aL}%
\left[ \left( \nu _{b}^{c}\right) _{L}\right] ^{c}-f_{ab}^{\nu }\frac{%
v_{s_{13}}}{\sqrt{2}}\overline{\nu }_{aR}^{c}\left( \nu _{bL}\right)
^{c}-f_{ab}^{\nu }\frac{\Lambda }{\sqrt{2}}\overline{\nu }_{aR}^{c}\left[
\left( \nu _{b}^{c}\right) _{L}\right] ^{c}+h.c. \\
&=&-f_{ab}^{\nu }\frac{v_{s_{11}}}{\sqrt{2}}\overline{\nu }_{aL}\nu
_{bL}^{c}-f_{ab}^{\nu }\frac{v_{s_{13}}}{\sqrt{2}}\overline{\nu }_{aL}\nu
_{bR}-f_{ab}^{\nu }\frac{v_{s_{13}}}{\sqrt{2}}\overline{\nu }_{aR}^{c}\nu
_{bL}^{c}-f_{ab}^{\nu }\frac{\Lambda }{\sqrt{2}}\overline{\nu }_{aR}^{c}\nu
_{bR}+h.c.
\end{eqnarray*}

We may arrange the mass terms as,

\begin{align*}
L_{mass}^{\nu S}& =-\left(
\begin{array}{cc}
\overline{\nu }_{aL} & \overline{\nu }_{aR}^{c}%
\end{array}%
\right) \frac{1}{\sqrt{2}}\left(
\begin{array}{cc}
f_{ab}^{\nu }v_{s_{1}} & f_{ab}^{\nu }v_{s_{3}} \\
f_{ab}^{\nu }v_{s_{3}} & f_{ab}^{\nu }\Lambda%
\end{array}%
\right) \left(
\begin{array}{c}
\nu _{bL}^{c} \\
\nu _{bR}%
\end{array}%
\right) +h.c. \\
& =-\left(
\begin{array}{cc}
\overline{\nu }_{aL} & \overline{\nu }_{aR}^{c}%
\end{array}%
\right) \frac{\sqrt{2}}{2}\left(
\begin{array}{cc}
v_{s_{1}}f_{ab}^{\nu } & v_{s_{3}}f_{ab}^{\nu } \\
v_{s_{3}}f_{ab}^{\nu } & \Lambda f_{ab}^{\nu }%
\end{array}%
\right) \left(
\begin{array}{c}
\nu _{bL}^{c} \\
\nu _{bR}%
\end{array}%
\right) +h.c. \\
& =-\frac{1}{2}\left(
\begin{array}{cc}
\overline{\nu }_{aL} & \overline{\nu }_{aR}^{c}%
\end{array}%
\right) \left(
\begin{array}{cc}
\sqrt{2}v_{s_{1}}f_{ab}^{\nu } & \sqrt{2}v_{s_{3}}f_{ab}^{\nu } \\
\sqrt{2}v_{s_{3}}f_{ab}^{\nu } & \sqrt{2}\Lambda f_{ab}^{\nu }%
\end{array}%
\right) \left(
\begin{array}{c}
\nu _{bL}^{c} \\
\nu _{bR}%
\end{array}%
\right) +h.c. \\
& =-\frac{1}{2}\left(
\begin{array}{cc}
\overline{\nu }_{aL} & \overline{\nu }_{aR}^{c}%
\end{array}%
\right) M_{\nu }\left(
\begin{array}{c}
\nu _{bL}^{c} \\
\nu _{bR}%
\end{array}%
\right) +h.c.
\end{align*}

With this result one can now easily understand Eq.\eqref{nuLagran}.

\subsection{Yukawa Interactions of Quarks}

We mentioned in the paper that we needed to invoke some discrete symmetries to prevent mixing between the SM and exotic quarks. We will explain this statement in more detail now.

The lepton number conserving terms in the renormalizable Yukawa Lagrangian
for the quarks sector are%
\begin{eqnarray}
L_{LNC} &=&h_{\alpha a}^{u}\overline{Q}_{\alpha L}\rho ^{\ast
}u_{aR}+h_{\alpha a}^{d}\overline{Q}_{\alpha L}\eta ^{\ast }d_{aR}+h^{U}%
\overline{Q}_{3L}\chi U_{R}  \notag \\
&&+h_{a}^{d}\overline{Q}_{3L}\rho d_{aR}+h_{a}^{u}\overline{Q}_{3L}\eta
u_{aR}+h_{\alpha \beta }^{D}\overline{Q}_{\alpha L}\chi ^{\ast }D_{\beta
R}+h.c.,  \label{Q1}
\end{eqnarray}%
while the lepton number violating terms of quarks are%
\begin{eqnarray}
L_{LNV} &=&s_{a}^{u}\overline{Q}_{3L}\chi u_{aR}+s_{\alpha a}^{d}\overline{Q}%
_{\alpha L}\chi ^{\ast }d_{\beta R}+s^{U}\overline{Q}_{3L}\eta U_{R} \\
&&+s_{\alpha a}^{D}\overline{Q}_{\alpha L}\eta ^{\ast }D_{aR}+s_{\alpha }^{D}%
\overline{Q}_{3L}\rho D_{\alpha R}+s_{\alpha }^{U}\overline{Q}_{\alpha
L}\rho ^{\ast }U_{R}+h.c.,
\label{eqpro}
\end{eqnarray}%
where $h$ and $s$ are constant couplings.

One might notice that the terms in Eq.\eqref{eqpro} will give rise to mass mixing terms involving the SM and exotic quarks, which can be problematic because they will lead to changes in the properties of the SM quarks. Therefore, one needs to prevent that and to do so invoke some discrete symmetries. The discrete symmetries have to be such that keep all the desired mass terms for the SM quarks and neutrino masses but forbid these ones. The set of discrete symmetries is, %
\begin{eqnarray*}
e_{aR} &\rightarrow &-e_{aR}, \\
u_{aR} &\rightarrow &-u_{aR}, \\
d_{aR} &\rightarrow &-d_{aR}, \\
U_{R} &\rightarrow &U_{R}, \\
D_{aR} &\rightarrow &D_{aR}, \\
\eta &\rightarrow &-\eta , \\
\rho &\rightarrow &-\rho , \\
\chi &\rightarrow &\chi ,
\end{eqnarray*}%
where $\alpha =1,2;$ $a=1,2,3.$

From Eq. (\eqref{Q1}), after spontaneous symmetry breaking we find,%
\begin{eqnarray}
L_{mass}^{Q} &=&\frac{v_{\rho }}{\sqrt{2}}h_{\alpha a}^{u}\overline{u}%
_{\alpha L}u_{aR}-\frac{v_{\eta }}{\sqrt{2}}h_{\alpha a}^{d}\overline{d}%
_{\alpha L}d_{aR}-\frac{v_{\eta }}{\sqrt{2}}h_{a}^{u}\overline{u}_{3L}u_{aR}
\notag \\
&&-\frac{v_{\rho }}{\sqrt{2}}h_{a}^{u}\overline{d}_{3L}d_{aR}-\frac{v_{\chi }%
}{\sqrt{2}}h_{U}\overline{U}_{L}U_{R}-\frac{v_{\chi }}{\sqrt{2}}h_{\alpha
\beta }^{D}\overline{D}_{\alpha L}D_{\beta R}+h.c..  \label{QM}
\end{eqnarray}

Thus, we can write the SM quark mass as follows,%
\begin{equation}
L_{mass}^{u}=-\frac{1}{\sqrt{2}}\left(
\begin{array}{ccc}
\overline{u}_{1L} & \overline{u}_{2L} & \overline{u}_{3L}%
\end{array}%
\right) \left(
\begin{array}{ccc}
-v_{\rho }h_{11}^{u} & -v_{\rho }h_{12}^{u} & -v_{\rho }h_{13}^{u} \\
-v_{\rho }h_{21}^{u} & -v_{\rho }h_{22}^{u} & -v_{\rho }h_{23}^{u} \\
v_{\eta }h_{31}^{u} & v_{\eta }h_{32}^{u} & v_{\eta }h_{33}^{u}%
\end{array}%
\right) \left(
\begin{array}{c}
u_{1R} \\
u_{2R} \\
u_{3R}%
\end{array}%
\right) +h.c.,
\end{equation}which leads to,
\begin{equation}
M_{u}=\frac{1}{\sqrt{2}}\left(
\begin{array}{ccc}
-v_{\rho }h_{11}^{u} & -v_{\rho }h_{12}^{u} & -v_{\rho }h_{13}^{u} \\
-v_{\rho }h_{21}^{u} & -v_{\rho }h_{22}^{u} & -v_{\rho }h_{23}^{u} \\
v_{\eta }h_{31}^{u} & v_{\eta }h_{32}^{u} & v_{\eta }h_{33}^{u}%
\end{array}%
\right) .
\end{equation}

Similarly, for the down-type quarks we get,

\begin{equation}
L_{mass}^{d}=-\frac{1}{\sqrt{2}}\left(
\begin{array}{ccc}
\overline{d}_{1L} & \overline{d}_{2L} & \overline{d}_{3L}%
\end{array}%
\right) \left(
\begin{array}{ccc}
v_{\eta }h_{11}^{d} & v_{\eta }h_{12}^{d} & v_{\eta }h_{13}^{d} \\
v_{\eta }h_{21}^{d} & v_{\eta }h_{22}^{d} & v_{\eta }h_{23}^{d} \\
v_{\rho }h_{31}^{d} & v_{\rho }h_{32}^{d} & v_{\rho }h_{33}^{d}%
\end{array}%
\right) \left(
\begin{array}{c}
d_{1R} \\
d_{2R} \\
d_{3R}%
\end{array}%
\right) +h.c.,
\end{equation}with,

\begin{equation}
M_{d}=\frac{1}{\sqrt{2}}\left(
\begin{array}{ccc}
v_{\eta }h_{11}^{d} & v_{\eta }h_{12}^{d} & v_{\eta }h_{13}^{d} \\
v_{\eta }h_{21}^{d} & v_{\eta }h_{22}^{d} & v_{\eta }h_{23}^{d} \\
v_{\rho }h_{31}^{d} & v_{\rho }h_{32}^{d} & v_{\rho }h_{33}^{d}%
\end{array}%
\right) .
\end{equation}%
We can see that the the type-up and type-down quarks are associated with the
VEVs $v_{\eta }$ and $v_{\rho },$ related to the electroweak scale.

From the Eq. (\ref{QM}), we can write the $U$ extra quark mass term%
\begin{equation}
M_{U}=\frac{1}{\sqrt{2}}v_{\chi }h_{U}.
\end{equation}

From the Eq. (\ref{QM}), we can write the $D$ extra quark mass lagrangian
\begin{equation*}
L_{mass}^{D}=-\frac{1}{\sqrt{2}}\left(
\begin{array}{cc}
\overline{D}_{1L} & \overline{D}_{2L}%
\end{array}%
\right) \left(
\begin{array}{cc}
v_{\chi }h_{11}^{D} & v_{\chi }h_{12}^{D} \\
v_{\chi }h_{21}^{D} & v_{\chi }h_{22}^{D}%
\end{array}%
\right) \left(
\begin{array}{c}
D_{1R} \\
D_{2R}%
\end{array}%
\right) +h.c..
\end{equation*}

The $D$ quark mass matrix in the basis ($D_{1},D_{2})$ is
\begin{equation*}
M_{D}=\frac{1}{\sqrt{2}}\left(
\begin{array}{cc}
v_{\chi }h_{11}^{D} & v_{\chi }h_{12}^{D} \\
v_{\chi }h_{21}^{D} & v_{\chi }h_{22}^{D}%
\end{array}%
\right) .
\end{equation*}%
Notice that the masses of $D$ quarks will depend the VEV $v_{\chi },$
related to the TeV-scale.

\subsection{Scalar Sector}

Considering the three scalar fields ($\chi ,$ $\eta ,$ $\rho ),$ the Higgs
potential more general, renormalizable and invariant on the $%
SU(3)_{L}\otimes U(1)_{X}$ symmetry group is
\begin{eqnarray}
V\left( \eta ,\chi ,\rho \right) &=&\mu _{\chi }^{2}\chi ^{\dagger }\chi
+\mu _{\eta }^{2}\eta ^{\dagger }\eta +\mu _{\rho }^{2}\rho ^{\dagger }\rho
\notag \\
&&+\lambda _{1}\left( \chi ^{\dagger }\chi \right) ^{2}+\lambda _{2}\left(
\eta ^{\dagger }\eta \right) ^{2}+\lambda _{3}\left( \rho ^{\dagger }\rho
\right) ^{2}  \notag \\
&&+\lambda _{4}\left( \chi ^{\dagger }\chi \right) \left( \eta ^{\dagger
}\eta \right) +\lambda _{5}\left( \chi ^{\dagger }\chi \right) \left( \rho
^{\dagger }\rho \right)  \notag \\
&&+\lambda _{6}\left( \eta ^{\dagger }\eta \right) \left( \rho ^{\dagger
}\rho \right) +\lambda _{7}\left( \chi ^{\dagger }\eta \right) \left( \eta
^{\dagger }\chi \right)  \notag \\
&&+\lambda _{8}\left( \chi ^{\dagger }\rho \right) \left( \rho ^{\dagger
}\chi \right) +\lambda _{9}\left( \eta ^{\dagger }\rho \right) \left( \rho
^{\dagger }\eta \right)  \notag \\
&&-\frac{f}{\sqrt{2}}\varepsilon ^{ijk}\eta _{i}\rho _{j}\chi _{k}+h.c.
\notag \\
&&+\mu _{4}^{2}\left( \chi ^{\dag }\eta +\eta ^{\dag }\chi \right) +\lambda
_{10}\left( \chi ^{\dagger }\chi \right) \left( \chi ^{\dag }\eta +\eta
^{\dag }\chi \right)  \notag \\
&&+\lambda _{11}\left( \eta ^{\dagger }\eta \right) \left( \chi ^{\dag }\eta
+\eta ^{\dag }\chi \right) +\lambda _{12}\left( \rho ^{\dagger }\rho \right)
\left( \chi ^{\dag }\eta +\eta ^{\dag }\chi \right)  \notag \\
&&+\lambda _{13}\left[ \left( \chi ^{\dag }\eta \right) \left( \chi ^{\dag
}\eta \right) +\left( \eta ^{\dag }\chi \right) \left( \eta ^{\dag }\chi
\right) \right]  \notag \\
&&+\lambda _{14}\left[ \left( \rho ^{\dag }\chi \right) \left( \eta ^{\dag
}\rho \right) +\left( \rho ^{\dag }\eta \right) \left( \chi ^{\dag }\eta
\right) \right] .
\end{eqnarray}%
where $\mu _{i}$ are constants, $\lambda _{i}$ and $f$ are constant
couplings. Furthermore, the additional terms of potential with sextet
scalar, as combinations with the others scalar triplets, are%
\begin{eqnarray}
&&V_{S}=\mu _{S}^{2}Tr\left( S^{\dagger }S\right) +\lambda _{10}Tr[\left(
S^{\dagger }S\right) ^{2}]+\lambda _{11}\left[ Tr\left( S^{\dagger }S\right) %
\right] ^{2}  \notag \\
&&+\left[ \lambda _{15}\left( \eta ^{\dagger }\eta \right) +\lambda
_{16}\left( \rho ^{\dagger }\rho \right) +\lambda _{17}\left( \chi ^{\dagger
}\chi \right) \right] Tr\left( S^{\dagger }S\right)  \notag \\
&&+\lambda _{18}\left( \chi ^{\dag }\eta +\eta ^{\dag }\chi \right) Tr\left(
S^{\dagger }S\right)  \notag \\
&&+\lambda _{S}\left( \varepsilon ^{ijk}\varepsilon ^{imn}\rho _{n}\rho
_{k}S_{li}S_{mj}+h.c.\right) +\lambda _{19}\left( \chi ^{\dagger }S\right)
\left( S^{\dagger }\chi \right)  \notag \\
&&+\lambda _{20}\left( \eta ^{\dagger }S\right) \left( S^{\dagger }\eta
\right) +\lambda _{21}\left( \rho ^{\dagger }S\right) \left( S^{\dagger
}\rho \right)  \notag \\
&&+\lambda _{22}(\varepsilon ^{ijk}\eta _{m}^{\ast }S_{mi}\rho _{j}\chi
_{k}+h.c.)+\lambda _{23}(\varepsilon ^{ijk}\chi _{m}^{\ast }S_{mi}\rho
_{j}\eta _{k}+h.c.)  \notag \\
&&+M_{1}\eta ^{T}S^{\dagger }\eta +M_{2}\chi ^{T}S^{\dagger }\chi +M_{3}\eta
^{\dagger }S\eta ^{\ast }+M_{4}\chi ^{\dagger }S\chi ^{\ast }.
\end{eqnarray}

The permitted terms of the scalar potential by the discrete symmetry are%
\begin{eqnarray}
V_{1}\left( \eta ,\chi ,\rho ,S\right) &=&\mu _{\chi }^{2}\chi ^{\dagger
}\chi +\mu _{\eta }^{2}\eta ^{\dagger }\eta +\mu _{\rho }^{2}\rho ^{\dagger
}\rho  \notag \\
&&+\lambda _{1}\left( \chi ^{\dagger }\chi \right) ^{2}+\lambda _{2}\left(
\eta ^{\dagger }\eta \right) ^{2}+\lambda _{3}\left( \rho ^{\dagger }\rho
\right) ^{2}  \notag \\
&&+\lambda _{4}\left( \chi ^{\dagger }\chi \right) \left( \eta ^{\dagger
}\eta \right) +\lambda _{5}\left( \chi ^{\dagger }\chi \right) \left( \rho
^{\dagger }\rho \right)  \notag \\
&&+\lambda _{6}\left( \eta ^{\dagger }\eta \right) \left( \rho ^{\dagger
}\rho \right) +\lambda _{7}\left( \chi ^{\dagger }\eta \right) \left( \eta
^{\dagger }\chi \right)  \notag \\
&&+\lambda _{8}\left( \chi ^{\dagger }\rho \right) \left( \rho ^{\dagger
}\chi \right) +\lambda _{9}\left( \eta ^{\dagger }\rho \right) \left( \rho
^{\dagger }\eta \right)  \notag \\
&&-\frac{f}{\sqrt{2}}\varepsilon ^{ijk}\eta _{i}\rho _{j}\chi _{k}+h.c.
\notag \\
&&+\lambda _{13}\left[ \left( \chi ^{\dag }\eta \right) \left( \chi ^{\dag
}\eta \right) +\left( \eta ^{\dag }\chi \right) \left( \eta ^{\dag }\chi
\right) \right]  \notag \\
&&+\mu _{S}^{2}Tr\left( S^{\dagger }S\right) +\lambda _{10}Tr[\left(
S^{\dagger }S\right) ^{2}]+\lambda _{11}\left[ Tr\left( S^{\dagger }S\right) %
\right] ^{2}  \notag \\
&&+\left[ \lambda _{15}\left( \eta ^{\dagger }\eta \right) +\lambda
_{16}\left( \rho ^{\dagger }\rho \right) +\lambda _{17}\left( \chi ^{\dagger
}\chi \right) \right] Tr\left( S^{\dagger }S\right)  \notag \\
&&+\lambda _{S}\left( \varepsilon ^{ijk}\varepsilon ^{imn}\rho _{n}\rho
_{k}S_{li}S_{mj}+h.c.\right) +\lambda _{19}\left( \chi ^{\dagger }S\right)
\left( S^{\dagger }\chi \right)  \notag \\
&&+\lambda _{20}\left( \eta ^{\dagger }S\right) \left( S^{\dagger }\eta
\right) +\lambda _{21}\left( \rho ^{\dagger }S\right) \left( S^{\dagger
}\rho \right)  \notag \\
&&+\lambda _{22}(\varepsilon ^{ijk}\eta _{m}^{\ast }S_{mi}\rho _{j}\chi
_{k}+h.c.)+\lambda _{23}(\varepsilon ^{ijk}\chi _{m}^{\ast }S_{mi}\rho
_{j}\eta _{k}+h.c.)  \notag \\
&&+M_{1}\eta ^{T}S^{\dagger }\eta +M_{2}\chi ^{T}S^{\dagger }\chi +M_{3}\eta
^{\dagger }S\eta ^{\ast }+M_{4}\chi ^{\dagger }S\chi ^{\ast }.
\end{eqnarray}

In this model, the scalar triplets develop non-trivial vacuum expectation
values (VEV), in order to engender correct spontaneous symmetry breaking
(SSB), as written below
\begin{equation}
\left\langle \chi \right\rangle =\frac{1}{\sqrt{2}}\left(
\begin{array}{c}
0 \\
0 \\
v_{\chi }%
\end{array}%
\right) ,\text{ }\left\langle \eta \right\rangle =\frac{1}{\sqrt{2}}\left(
\begin{array}{c}
v_{\eta } \\
0 \\
0%
\end{array}%
\right) ,\text{ }\left\langle \rho \right\rangle =\frac{1}{\sqrt{2}}\left(
\begin{array}{c}
0 \\
v_{\rho } \\
0%
\end{array}%
\right) .\text{ }
\end{equation}%
Observe that the scalar triplet $\chi $ develops VEV only on the third
neutral component, while $\eta $ develops VEV only on the first neutral
component. The steps of symmetry breaking transition is given by $%
SU(3)_{L}\otimes U(1)_{X}\overset{\left\langle \chi \right\rangle }{%
\rightarrow }SU(2)_{L}\otimes U(1)_{Y},\ $while$\ \ SU(2)_{L}\otimes U(1)_{Y}%
\overset{\left\langle \eta \right\rangle ,\left\langle \rho \right\rangle }{%
\rightarrow }U(1)_{Q},$ i.e., the triplet $\chi $ develops VEV breaking the $%
3-3-1$ gauge symmetry to SM, while the triplets $\eta $ and $\rho $ develop
VEV breaking the SM gauge symmetry to the QED (Quantum Electrodynamics).

Regarding the scalar sextet, the three neutral components develop VEVs in
the following way:
\begin{equation}
\left\langle S\right\rangle =\frac{1}{\sqrt{2}}\left(
\begin{tabular}{lll}
$v_{s_{1}}$ & $0$ & $v_{s_{3}}$ \\
$0$ & $0$ & $0$ \\
$v_{s_{3}}$ & $0$ & $\Lambda $%
\end{tabular}%
\right)
\end{equation}%
We will see that $v_{s_{1},}$ $v_{s_{3}}$ and $\Lambda $ are responsible for
the mass for the left-handed neutrinos, while $\Lambda $ is responsible for
the right-handed neutrinos Dirac masses. After the SSB of $SU(3)_{L}\otimes
U(1)_{X}$ to $SU(2)_{L}\otimes U(1)_{Y}$, the scalar sextet results in a
triplet plus a doublet and a singlet ($\mathbf{6}\rightarrow \mathbf{3}+%
\mathbf{2}+\mathbf{1})$, as follows
\begin{equation}
S\rightarrow S_{1(\mathbf{1},\mathbf{3},-2)}+S_{2(\mathbf{1},\mathbf{2}%
,-1)}+S_{3(\mathbf{1},\mathbf{1},0)},
\end{equation}%
where
\begin{equation}
S_{1}=\left(
\begin{tabular}{ll}
$S_{11}^{0}$ & $S_{12}^{-}$ \\
$S_{12}^{-}$ & $S_{22}^{--}$%
\end{tabular}%
\right) ,\text{ }S_{2}=\left(
\begin{tabular}{l}
$S_{13}^{0}$ \\
$S_{23}^{-}$%
\end{tabular}%
\right) ,\text{ }S_{3}=S_{33}^{0}.
\end{equation}

In this model, the lepton number distribution of the scalars is%
\begin{eqnarray}
L(\eta _{3}^{0},S_{33}^{0},\rho _{3}^{+}) &=&-2, \\
L(\chi _{1}^{0},\chi _{2}^{-},S_{11}^{0},S_{12}^{-},S_{22}^{--}) &=&+2.
\end{eqnarray}%
We can see that $S_{12}^{-}$ and $S_{22}^{--}$ carry two lepton numbers.
Both are essential in our analysis of charged lepton flavor violating decay
of muon.

\subsection{Charged Lepton Flavor Interactions}

\indent

The Yukawa Lagrangian with charged lepton flavor violation (CLFV) is given by
\begin{equation}
L_{CLFV}\supset f_{ab}^{\nu }(\overline{\psi }_{aL}^{c})_{m}(\psi
_{bL})_{n}(S^{\ast })_{mn}+h.c.,  \label{L}
\end{equation}%
where $a,b=1,2,3$ indicates the lepton generations and $m,n=1,2,3$ indicates
the entries of the sextet, $f_{ab}$ is symmetric. The CLFV interactions in the $%
\mu -e$ sector results from taking $ a = 1 , b = 2 $ in the above equation,
\begin{equation}
f_{ab}\overline{\left( \psi _{aL}\right) _{m}^{c}}\left( \psi _{bL}\right)
_{n}S_{mn}^{\ast }\overset{a=1,b=2}{\rightarrow }f_{12}\overline{\left(
e_{L}\right) ^{c}}\left( \mu _{L}\right) S^{++}  \notag .
\end{equation}%

Using the following relations:%
\begin{align*}
\left( \Psi ^{c}\right) _{L}& =\left( \Psi _{R}\right) ^{c} \\
\left( \Psi ^{c}\right) _{R}& =\left( \Psi _{L}\right) ^{c} \\
\left( \overline{\Psi }^{c}\right) _{L}& =\left( \overline{\Psi }_{R}\right)
^{c} \\
\left( \overline{\Psi }^{c}\right) _{R}& =\left( \overline{\Psi }_{L}\right)
^{c}.
\end{align*}
we have,
\begin{align}
f_{12} \overline{\left( e_{L}\right) ^{c}}\left( \mu _{L}\right) S^{++}&
=f_{12}\overline{\left( e^{c}\right) _{R}}\left( \mu _{L}\right) S^{++}
\notag \\
& =f_{12}\left[ \left( e^{c}\right) _{R}\right] ^{\dag }\gamma ^{0}\left(
\mu _{L}\right) S^{++}  \notag \\
& =f_{12}\left[ \frac{1}{2}\left( 1+\gamma _{5}\right) e^{c}\right] ^{\dag
}\gamma ^{0}\left( \mu _{L}\right) S^{++}  \notag \\
& =f_{12}\left[ \frac{1}{2}\left( 1+\gamma _{5}\right) e^{c}\right] ^{\dag
}\gamma ^{0}\frac{1}{2}\left( 1-\gamma _{5}\right) \mu S^{++}  \notag \\
& =f_{12}\left( e^{c\dag }\right) \frac{1}{2}\left( 1-\gamma _{5}\right)
\gamma ^{0}\frac{1}{2}\left( 1-\gamma _{5}\right) \mu S^{++}  \notag \\
& =f_{12}\left( e^{c\dag }\gamma ^{0}\right) \frac{1}{2}\left( 1-\gamma
_{5}\right) \mu S^{++}  \notag \\
& =f_{12}\overline{e^{c}}\frac{1}{2}\left( 1-\gamma _{5}\right) \mu S^{++}
\notag \\
f_{12} \overline{\left( e_{L}\right) ^{c}}\left( \mu _{L}\right) S^{++}& =\frac{1}{2}f_{12}\overline{e^{c}}\mu S^{++}-\frac{1}{2}f_{12}\overline{%
e^{c}}\gamma _{5}\mu S^{++} .
\end{align}

We see that these terms directly induce $\mu \rightarrow e\gamma $ lepton flavor
violating decay mediated by the doubly charged scalar $S^{++}$. Proceeding in an analogous way, we can easily obtain the relevant CLFV terms mediated by the singly charged scalar $ S ^+ $.


\section{Derivation of the $\MEG$ decay }

The on-shell amplitude for the $l_{j}^{-}\rightarrow l_{i}^{-}\gamma $ process is
written in the form%
\begin{equation}
\mathcal{M=}e\varepsilon _{\mu }^{\ast }\overline{u}_{i}\left( p+q\right) %
\left[ m_{l_{j}}i\sigma ^{\mu \nu }q_{\nu }\left(
A_{2}^{L}P_{L}+A_{2}^{R}P_{R}\right) \right] u_{j}\left( p\right) ,
\end{equation}%
where $\varepsilon ^{\mu }$ is the polarization vector of the on-shell photon, $e$ is
the electric charge, $u_{i}$ is the spinor of $l_{i}^{-},$ $u_{j}$ is the
spinor of $l_{j}^{-},$ $m_{l_{j}}$ is the mass of $l_{j}$ lepton, $\sigma
^{\mu \nu }$ is the comutator of $\gamma $ matrices, $q$ is the photon
4-momentum, $p$ is the $l_{j}$ lepton 4-momentum, $P_{L}$ is the chirality
projector of the left-handed lepton, $P_{R}$ is the chirality projection of
the right-handed lepton, $A_{2}^{L}$ and $A_{2}^{R}$ are coupling constants.

The squared modulus of the amplitude is%
\begin{eqnarray}
\left\vert \mathcal{M}\right\vert ^{2} &\mathcal{=}&e^{2}\left\{ \varepsilon
_{\mu }^{\ast }\overline{u}_{i}\left( p+q\right) \left[ m_{l_{j}}i\sigma
^{\mu \nu }q_{\nu }\left( A_{2}^{L}P_{L}+A_{2}^{R}P_{R}\right) \right]
u_{j}\left( p\right) \right\} ^{\dagger }\times  \notag \\
&&\left\{ \varepsilon _{\alpha }^{\ast }\overline{u}_{i}\left( p+q\right) %
\left[ m_{l_{j}}i\sigma ^{\alpha \beta }q_{\beta }\left(
A_{2}^{L}P_{L}+A_{2}^{R}P_{R}\right) \right] u_{j}\left( p\right) \right\} ,
\\
&&  \notag \\
&=&e^{2}m_{l_{j}}^{2}\left\{ \varepsilon _{\mu }\overline{u}_{i}\left(
p+q\right) \left[ \sigma ^{\mu \nu }q_{\nu }\left(
A_{2}^{L}P_{L}+A_{2}^{R}P_{R}\right) \right] u_{j}\left( p\right) \right\}
^{\dagger }\times  \notag \\
&&\left\{ \varepsilon _{\alpha }\overline{u}_{i}\left( p+q\right) \left[
\sigma ^{\alpha \beta }q_{\beta }\left( A_{2}^{L}P_{L}+A_{2}^{R}P_{R}\right) %
\right] u_{j}\left( p\right) \right\} .
\end{eqnarray}%
Summing over the polarization states of the on-shell photon, 
we have the squared amplitude written in the form%
\begin{eqnarray}
\sum\limits_{\lambda }\left\vert \mathcal{M}\right\vert ^{2}
&=&e^{2}m_{l_{j}}^{2}\left( \sum_{\lambda }\varepsilon _{\mu }^{\lambda
}\varepsilon _{\alpha }^{\ast \lambda }\right) \left\{ \overline{u}%
_{i}\left( p+q\right) \left[ \sigma ^{\mu \nu }q_{\nu }\left(
A_{2}^{L}P_{L}+A_{2}^{R}P_{R}\right) \right] u_{j}\left( p\right) \right\}
^{\dagger }\times  \notag \\
&&\left\{ \overline{u}_{m}\left( p+q\right) \left[ \sigma ^{\alpha \beta
}q_{\beta }\left( A_{2}^{L}P_{L}+A_{2}^{R}P_{R}\right) \right] u_{n}\left(
p\right) \right\} .
\end{eqnarray}%
Using the completeness relation $\sum_{\lambda }\varepsilon _{\mu }^{\lambda
}\varepsilon _{\alpha }^{\ast \lambda }=-g_{\mu \alpha }$
, we get%
\begin{eqnarray}
\sum\limits_{\lambda }\left\vert \mathcal{M}\right\vert ^{2}
&=&-e^{2}m_{l_{j}}^{2}g_{\mu \alpha }\left\{ \overline{u}_{i}\left(
p+q\right) \left[ \sigma ^{\mu \nu }q_{\nu }\left(
A_{2}^{L}P_{L}+A_{2}^{R}P_{R}\right) \right] u_{j}\left( p\right) \right\}
^{\dagger }\times  \notag \\
&&\left\{ \overline{u}_{m}\left( p+q\right) \left[ \sigma ^{\alpha \beta
}q_{\beta }\left( A_{2}^{L}P_{L}+A_{2}^{R}P_{R}\right) \right] u_{n}\left(
p\right) \right\} .
\end{eqnarray}

In order to obtain the squared amplitude explicitly, we need to calculate the
term%
\begin{equation}
\left[ \overline{u}_{i}\Gamma ^{\mu }u_{j}\right] ^{\dagger }=u_{j}^{\dagger
}\Gamma ^{\dagger \mu }\gamma ^{0}u_{i},  
\end{equation}%
where%
\begin{equation}
\label{cap_gamma_dagger}
\Gamma ^{\dagger \mu }=\left( A_{2}^{L}P_{L}+A_{2}^{R}P_{R}\right) ^{\dagger
}q_{\nu }\sigma ^{\dagger \mu \nu },
\end{equation}%
where $P_{L}=\left( 1-\gamma _{5}\right) /2$ and $P_{R}=\left( 1+\gamma
_{5}\right) /2$ are the chirality projectors$.$

By using the following expression
\begin{eqnarray}
\sigma ^{\dagger \mu \nu } &=&\frac{i}{2}\gamma ^{0}\left[ \gamma ^{\mu
},\gamma ^{\nu }\right] \gamma ^{0},  \notag \\
&=&\gamma ^{0}\sigma ^{\mu \nu }\gamma ^{0},
\end{eqnarray}%
the Eq. (\ref{cap_gamma_dagger}) can be written as%
\begin{eqnarray}
\Gamma ^{\dagger \mu } &=&\left( A_{2}^{L}P_{L}+A_{2}^{R}P_{R}\right)
^{\dagger }q_{\nu }\gamma ^{0}\sigma ^{\mu \nu }\gamma ^{0},  \notag \\
&=&\left( A_{2}^{L\ast }P_{L}^{\dagger }+A_{2}^{R\ast }P_{R}^{\dagger
}\right) q_{\nu }\gamma ^{0}\sigma ^{\mu \nu }\gamma ^{0}.  
\end{eqnarray}

Reminding that $P_{L}^{\dagger }=P_{L}$ and $P_{R}^{\dagger }=P_{R},$ so
we get%
\begin{eqnarray}
\Gamma ^{\dagger \mu } &=&\left( A_{2}^{L\ast }P_{L}+A_{2}^{R\ast
}P_{R}\right) q_{\nu }\sigma ^{\mu \nu }\gamma ^{0}\gamma ^{0},  \notag \\
&=&\left( A_{2}^{L\ast }P_{L}+A_{2}^{R\ast }P_{R}\right) q_{\nu }\sigma
^{\mu \nu }.
\end{eqnarray}%
Then, we have the following result%
\begin{eqnarray}
\overline{u}_{j}\Gamma ^{\dagger \mu }u_{i} &=&u_{j}^{\dagger }\left(
A_{2}^{L\ast }P_{L}+A_{2}^{R\ast }P_{R}\right) q_{\nu }\sigma ^{\mu \nu
}\gamma ^{0}u_{i},  \notag \\
&=&u_{j}^{\dagger }\left( A_{2}^{L\ast }P_{L}+A_{2}^{R\ast }P_{R}\right)
q_{\nu }\gamma ^{0}\sigma ^{\mu \nu }u_{i}.  
\end{eqnarray}

Given that $P_{L}\gamma ^{0}=\gamma ^{0}P_{R}$ and $P_{R}\gamma ^{0}=\gamma
^{0}P_{L},$ we have%
\begin{eqnarray}
\overline{u}_{j}\Gamma ^{\dagger \mu }u_{i} &=&u_{j}^{\dagger }\gamma
^{0}\left( A_{2}^{L\ast }P_{R}+A_{2}^{R\ast }P_{L}\right) q_{\nu }q^{\nu
}\sigma ^{\mu \nu }u_{i},  \notag \\
&=&\overline{u}_{j}\left( A_{2}^{L\ast }P_{R}+A_{2}^{R\ast }P_{L}\right)
q_{\nu }\sigma ^{\mu \nu }u_{i}.
\end{eqnarray}

Therefore the squared amplitude is written in the following form%
\begin{equation}
\sum\limits_{\lambda }\left\vert \mathcal{M}\right\vert
^{2}=-e^{2}m_{l_{j}}^{2}g_{\mu \alpha }\left\{ \overline{u}_{j}\left[ \left(
A_{2}^{L\ast }P_{R}+A_{2}^{R\ast }P_{L}\right) q_{\nu }\sigma ^{\mu \nu }%
\right] u_{i}\right\} \left\{ \overline{u}_{n}\left( \sigma ^{\alpha \beta
}q_{\beta }\left( A_{2}^{L}P_{L}+A_{2}^{R}P_{R}\right) \right) u_{m}\right\},
\end{equation}%
Summing over the spins of the fermions and writing the multiplication matrix in index form, we get%
\begin{eqnarray}
\sum\limits_{spins}\sum\limits_{\lambda }\left\vert \mathcal{M}\right\vert
^{2} &=&-e^{2}m_{l_{j}}^{2}g_{\mu \alpha }\sum\limits_{r,s}\left\{
\overline{u}_{j}^{r}\left[ \left( A_{2}^{L\ast }P_{R}+A_{2}^{R\ast
}P_{L}\right) q_{\nu }\sigma ^{\mu \nu }\right] _{ji}u_{i}^{s}\right\} \times
\notag \\
&&\left\{ \overline{u}_{n}^{s}\left[ \sigma ^{\alpha \beta }q_{\beta }\left(
A_{2}^{L}P_{L}+A_{2}^{R}P_{R}\right) \right] _{nm}u_{m}^{r}\right\} , \\
&&  \notag \\
&=&-e^{2}m_{l_{j}}^{2}g_{\mu \alpha }\left( \sum\limits_{r}u_{m}^{r}%
\overline{u}_{j}^{r}\right) \left( \sum\limits_{s}u_{i}^{s}\overline{u}%
_{n}^{s}\right) \left\{ \left[ \left( A_{2}^{L\ast }P_{R}+A_{2}^{R\ast
}P_{L}\right) q_{\nu }\sigma ^{\mu \nu }\right] _{ji}\right\} \times  \notag
\\
&&\left\{ \left[ \sigma ^{\alpha \beta }q_{\beta }\left(
A_{2}^{L}P_{L}+A_{2}^{R}P_{R}\right) \right] _{nm}\right\} .
\end{eqnarray}
Using the completeness relation
\begin{eqnarray}
\sum\limits_{a}u^{a}%
\overline{u}^{a}=\rlap{\hbox{$\mskip
1 mu /$}}p+m , 
\end{eqnarray}
and doing the average of the initial spin of fermion, we get%
\begin{eqnarray}
\frac{1}{2}\sum\limits_{spins}\sum\limits_{\lambda }\left\vert \mathcal{M}%
\right\vert ^{2} &=&-\frac{1}{2}e^{2}m_{l_{j}}^{2}g_{\mu \alpha }\left( {%
\rlap{\hbox{$\mskip
1 mu /$}}}p_{1}+{\rlap{\hbox{$\mskip
1 mu /$}}}q+m_{1}\right) _{mj}\left( {\rlap{\hbox{$\mskip
1 mu /$}}}p_{2}+m_{2}\right) _{in}\left\{ \left[ \left( A_{2}^{L\ast
}P_{R}+A_{2}^{R\ast }P_{L}\right) q_{\nu }\sigma ^{\mu \nu }\right]
_{ji}\right\} \times  \notag \\
&&\left\{ \left[ \sigma ^{\alpha \beta }q_{\beta }\left(
A_{2}^{L}P_{L}+A_{2}^{R}P_{R}\right) \right] _{nm}\right\} .
\end{eqnarray}

We need to put back the equation above into normal matrix multiplication order
\begin{eqnarray}
\frac{1}{2}\sum\limits_{spins}\sum\limits_{\lambda }\left\vert \mathcal{M}%
\right\vert ^{2}
&=&-\frac{1}{2}e^{2}m_{l_{j}}^{2}g_{\mu \alpha }\left( {%
\rlap{\hbox{$\mskip
1 mu /$}}}p_{1}+{\rlap{\hbox{$\mskip
1 mu /$}}}q+m_{1}\right) _{mj}\left\{ \left[ \left( A_{2}^{L\ast
}P_{R}+A_{2}^{R\ast }P_{L}\right) q_{\nu }\sigma ^{\mu \nu }\right]
_{ji}\right\} \left( {\rlap{\hbox{$\mskip
1 mu /$}}}p_{2}+m_{2}\right) _{in}\times  \notag \\
&&\left\{ \left[ \sigma ^{\alpha \beta }q_{\beta }\left(
A_{2}^{L}P_{L}+A_{2}^{R}P_{R}\right) \right] _{nm}\right\} .
\end{eqnarray}

We can write the equation above in terms of a trace
\begin{eqnarray}
\frac{1}{2}\sum\limits_{spins}\sum\limits_{\lambda }\left\vert \mathcal{M}%
\right\vert ^{2}
&=&-\frac{1}{2}e^{2}m_{l_{j}}^{2}g_{\mu \alpha }q_{\nu }q_{\beta }\text{Tr}\{\left[
\left( {\rlap{\hbox{$\mskip
1 mu /$}}}p_{1}+{\rlap{\hbox{$\mskip
1 mu /$}}}q+m_{1}\right) \left( A_{2}^{L\ast }P_{R}+A_{2}^{R\ast
}P_{L}\right) \sigma ^{\mu \nu }\left( {\rlap{\hbox{$\mskip
1 mu /$}}}p_{2}+m_{2}\right) \right] \times  \notag \\
&&\left[ \sigma ^{\alpha \beta }\left( A_{2}^{L}P_{L}+A_{2}^{R}P_{R}\right) %
\right] \}.
\end{eqnarray}%
Using $\left\{ \gamma _{5},\gamma ^{\mu }\right\} =0$ and $P_{L}^{2}=P_{L},$
$P_{R}^{2}=P_{R},$ $P_{L}P_{R}=P_{R}P_{L}=0,$ we have%
\begin{equation}
\frac{1}{2}\sum\limits_{spins}\sum\limits_{\lambda }\left\vert \mathcal{M}%
\right\vert ^{2}=-\frac{e^{2}}{2}m_{l_{j}}^{2}g_{\mu \alpha }q_{\nu
}q_{\beta }\text{ }\text{Tr}\left[ \left( {\rlap{\hbox{$\mskip
1 mu /$}}}p_{1}+{\rlap{\hbox{$\mskip
1 mu /$}}}q+m_{1}\right) \left( \left\vert A_{2}^{L}\right\vert
^{2}P_{R}+\left\vert A_{2}^{R}\right\vert ^{2}P_{L}\right) \sigma ^{\mu \nu
}\left( {\rlap{\hbox{$\mskip
1 mu /$}}}p_{2}+m_{2}\right) \sigma ^{\alpha \beta }\right] .
\end{equation}

Let's calculate, ignoring $m_{1},$ the following term%
\begin{eqnarray}
\left( {\rlap{\hbox{$\mskip
1 mu /$}}}p_{1}+{\rlap{\hbox{$\mskip
1 mu /$}}}q\right) \left( \left\vert A_{2}^{L}\right\vert
^{2}P_{R}+\left\vert A_{2}^{R}\right\vert ^{2}P_{L}\right) &=&\left( {%
\rlap{\hbox{$\mskip
1 mu /$}}}p_{1}+{\rlap{\hbox{$\mskip
1 mu /$}}}q\right) \left\vert A_{2}^{L}\right\vert ^{2}\frac{\left( 1+\gamma
_{5}\right) }{2}+\left( {\rlap{\hbox{$\mskip
1 mu /$}}}p_{1}+{\rlap{\hbox{$\mskip
1 mu /$}}}q\right) \left\vert A_{2}^{R}\right\vert ^{2}\frac{\left( 1-\gamma
_{5}\right) }{2}  \notag \\
&=&\frac{1}{2}\left( {\rlap{\hbox{$\mskip
1 mu /$}}}p_{1}\left\vert A_{2}^{L}\right\vert ^{2}+{%
\rlap{\hbox{$\mskip
1 mu /$}}}p_{1}\left\vert A_{2}^{L}\right\vert ^{2}\gamma _{5}+{%
\rlap{\hbox{$\mskip
1 mu /$}}}p_{1}\left\vert A_{2}^{R}\right\vert ^{2}-{%
\rlap{\hbox{$\mskip
1 mu /$}}}p_{1}\left\vert A_{2}^{R}\right\vert ^{2}\gamma _{5}\right)+  \notag
\\
&&\frac{1}{2}\left( {\rlap{\hbox{$\mskip
1 mu /$}}}q\left\vert A_{2}^{L}\right\vert ^{2}+{%
\rlap{\hbox{$\mskip
1 mu /$}}}q\left\vert A_{2}^{L}\right\vert ^{2}\gamma _{5}+{%
\rlap{\hbox{$\mskip
1 mu /$}}}q\left\vert A_{2}^{R}\right\vert ^{2}-{%
\rlap{\hbox{$\mskip
1 mu /$}}}q\left\vert A_{2}^{R}\right\vert ^{2}\gamma _{5}\right) .
\end{eqnarray}

Now we need to calculate this equation below%
\begin{eqnarray}
\sigma ^{\mu \nu }\left( {\rlap{\hbox{$\mskip
1 mu /$}}}p_{2}+m_{2}\right) \sigma ^{\alpha \beta } &=&\frac{i}{2}\left[
\gamma ^{\mu },\gamma ^{\nu }\right] \left( {\rlap{\hbox{$\mskip
1 mu /$}}}p_{2}+m_{2}\right) \frac{i}{2}\left[ \gamma ^{\alpha },\gamma
^{\beta }\right] ,  \notag \\
&=&-\frac{1}{4}\left( \gamma ^{\mu }\gamma ^{\nu }-\gamma ^{\nu }\gamma
^{\mu }\right) \left( {\rlap{\hbox{$\mskip
1 mu /$}}}p_{2}+m_{2}\right) \left( \gamma ^{\alpha }\gamma ^{\beta }-\gamma
^{\beta }\gamma ^{\alpha }\right) .
\end{eqnarray}

Let's calculate explicitly the equation above:%
\begin{eqnarray}
&&\left( \gamma ^{\mu }\gamma ^{\nu }-\gamma ^{\nu }\gamma ^{\mu }\right)
\left( {\rlap{\hbox{$\mskip
1 mu /$}}}p_{2}+m_{2}\right) \left( \gamma ^{\alpha }\gamma ^{\beta }-\gamma
^{\beta }\gamma ^{\alpha }\right)   \notag \\
&=&\gamma ^{\mu }\gamma ^{\nu }\rlap{\hbox{$\mskip
1 mu /$}}p_{2}\gamma ^{\alpha }\gamma ^{\beta
}+m_{2}\gamma ^{\mu }\gamma ^{\nu }\gamma ^{\alpha }\gamma ^{\beta }-\gamma
^{\nu }\gamma ^{\mu }\rlap{\hbox{$\mskip
1 mu /$}}p_{2}\gamma ^{\alpha }\gamma ^{\beta }  \notag \\
&&-m_{2}\gamma ^{\nu }\gamma ^{\mu }\gamma ^{\alpha }\gamma ^{\beta }-\gamma
^{\mu }\gamma ^{\nu }\rlap{\hbox{$\mskip
1 mu /$}}p_{2}\gamma ^{\beta }\gamma ^{\alpha
}-m_{2}\gamma ^{\mu }\gamma ^{\nu }\gamma ^{\beta }\gamma ^{\alpha }  \notag
\\
&&+\gamma ^{\nu }\gamma ^{\mu }\rlap{\hbox{$\mskip
1 mu /$}}p_{2}\gamma ^{\beta }\gamma ^{\alpha
}+m_{2}\gamma ^{\nu }\gamma ^{\mu }\gamma ^{\beta }\gamma ^{\alpha }.  
\end{eqnarray}

Inserting the $g_{\mu \alpha }$,
\begin{eqnarray}
&&g_{\mu \alpha }\left( \gamma ^{\mu }\gamma ^{\nu }-\gamma ^{\nu }\gamma
^{\mu }\right) \left( {\rlap{\hbox{$\mskip
1 mu /$}}}p_{2}+m_{2}\right) \left( \gamma ^{\alpha }\gamma ^{\beta }-\gamma
^{\beta }\gamma ^{\alpha }\right)   \notag \\
&=&\gamma _{\alpha }\gamma ^{\nu }\rlap{\hbox{$\mskip
1 mu /$}}p_{2}\gamma ^{\alpha }\gamma ^{\beta
}+m_{2}\gamma _{\alpha }\gamma ^{\nu }\gamma ^{\alpha }\gamma ^{\beta
}-\gamma ^{\nu }\gamma _{\alpha }\rlap{\hbox{$\mskip
1 mu /$}}p_{2}\gamma ^{\alpha }\gamma ^{\beta }
\notag \\
&&-m_{2}\gamma ^{\nu }\gamma _{\alpha }\gamma ^{\alpha }\gamma ^{\beta
}-\gamma _{\alpha }\gamma ^{\nu }\rlap{\hbox{$\mskip
1 mu /$}}p_{2}\gamma ^{\beta }\gamma ^{\alpha
}-m_{2}\gamma _{\alpha }\gamma ^{\nu }\gamma ^{\beta }\gamma ^{\alpha }
\notag \\
&&+\gamma ^{\nu }\gamma _{\alpha }\rlap{\hbox{$\mskip
1 mu /$}}p_{2}\gamma ^{\beta }\gamma ^{\alpha
}+m_{2}\gamma ^{\nu }\gamma _{\alpha }\gamma ^{\beta }\gamma ^{\alpha } \notag
\end{eqnarray}%
\begin{eqnarray}
=&&\gamma _{\alpha }\gamma ^{\nu }\rlap{\hbox{$\mskip
1 mu /$}}p_{2}\gamma ^{\alpha }\gamma ^{\beta
}+m_{2}\gamma _{\alpha }\gamma ^{\nu }\gamma ^{\alpha }\gamma ^{\beta
}-\gamma ^{\nu }\gamma _{\alpha }\rlap{\hbox{$\mskip
1 mu /$}}p_{2}\gamma ^{\alpha }\gamma ^{\beta }
\notag \\
&&-m_{2}\gamma ^{\nu }\gamma _{\alpha }\gamma ^{\alpha }\gamma ^{\beta
}-\gamma _{\alpha }\gamma ^{\nu }\rlap{\hbox{$\mskip
1 mu /$}}p_{2}\gamma ^{\beta }\gamma ^{\alpha
}-m_{2}\gamma _{\alpha }\gamma ^{\nu }\gamma ^{\beta }\gamma ^{\alpha }
\notag \\
&&+\gamma ^{\nu }\gamma _{\alpha }\rlap{\hbox{$\mskip
1 mu /$}}p_{2}\gamma ^{\beta }\gamma ^{\alpha
}+m_{2}\gamma ^{\nu }\gamma _{\alpha }\gamma ^{\beta }\gamma ^{\alpha }.
\end{eqnarray}

Writing $\rlap{\hbox{$\mskip
1 mu /$}}p_{2}=p_{2\eta }\gamma ^{\eta },$ we get%
\begin{eqnarray}
&=&p_{2\eta }\left( \gamma _{\alpha }\gamma ^{\nu }\gamma ^{\eta }\gamma
^{\alpha }\gamma ^{\beta }-\gamma ^{\nu }\gamma _{\alpha }\gamma ^{\eta
}\gamma ^{\alpha }\gamma ^{\beta }-\gamma _{\alpha }\gamma ^{\nu }\gamma
^{\eta }\gamma ^{\beta }\gamma ^{\alpha }+\gamma ^{\nu }\gamma _{\alpha
}\gamma ^{\eta }\gamma ^{\beta }\gamma ^{\alpha }\right)   \notag \\
&&+m_{2}\left( \gamma _{\alpha }\gamma ^{\nu }\gamma ^{\alpha }\gamma
^{\beta }-\gamma ^{\nu }\gamma _{\alpha }\gamma ^{\alpha }\gamma ^{\beta
}-\gamma _{\alpha }\gamma ^{\nu }\gamma ^{\beta }\gamma ^{\alpha }+\gamma
^{\nu }\gamma _{\alpha }\gamma ^{\beta }\gamma ^{\alpha }\right) .
\end{eqnarray}

Let's simplify the term above, using the contraction identities%
\begin{eqnarray}
\gamma _{\alpha }\gamma ^{\alpha } &=&4, \\
\gamma _{\alpha }\gamma ^{\mu }\gamma ^{\alpha } &=&-2\gamma ^{\mu }, \\
\gamma _{\alpha }\gamma ^{\mu }\gamma ^{\alpha } &=&-2\gamma ^{\mu }, \\
\gamma _{\alpha }\gamma ^{\mu }\gamma ^{\nu }\gamma ^{\alpha } &=&4g^{\mu
\nu }, \\
\gamma _{\alpha }\gamma ^{\mu }\gamma ^{\nu }\gamma ^{\kappa }\gamma
^{\alpha } &=&-2\gamma ^{\mu }\gamma ^{\nu }\gamma ^{\kappa },
\end{eqnarray}%
It leads us to the result%
\begin{eqnarray}
&&p_{2\eta }\left( \gamma _{\alpha }\gamma ^{\nu }\gamma ^{\eta }\gamma
^{\alpha }\gamma ^{\beta }-\gamma ^{\nu }\gamma _{\alpha }\gamma ^{\eta
}\gamma ^{\alpha }\gamma ^{\beta }-\gamma _{\alpha }\gamma ^{\nu }\gamma
^{\eta }\gamma ^{\beta }\gamma ^{\alpha }+\gamma ^{\nu }\gamma _{\alpha
}\gamma ^{\eta }\gamma ^{\beta }\gamma ^{\alpha }\right)   \notag \\
&&+m_{2}\left( \gamma _{\alpha }\gamma ^{\nu }\gamma ^{\alpha }\gamma
^{\beta }-\gamma ^{\nu }\gamma _{\alpha }\gamma ^{\alpha }\gamma ^{\beta
}-\gamma _{\alpha }\gamma ^{\nu }\gamma ^{\beta }\gamma ^{\alpha }+\gamma
^{\nu }\gamma _{\alpha }\gamma ^{\beta }\gamma ^{\alpha }\right)   \notag \\
&=&p_{2\eta }\left( 4g^{\nu \eta }\gamma ^{\beta }+2\gamma ^{\nu }\gamma
^{\eta }\gamma ^{\beta }+2\gamma ^{\nu }\gamma ^{\eta }\gamma ^{\beta
}+4\gamma ^{\nu }g_{\eta \beta }\right)   \notag \\
&&+m_{2}\left( -2\gamma ^{\nu }\gamma ^{\beta }-4\gamma ^{\nu }\gamma
^{\beta }-4g^{\nu \beta }-2\gamma ^{\nu }\gamma ^{\beta }\right) ,  \notag \\
&=&p_{2\eta }\left( 4g^{\nu \eta }\gamma ^{\beta }+4\gamma ^{\nu }\gamma
^{\eta }\gamma ^{\beta }+4\gamma ^{\nu }g^{\eta \beta }\right) +m_{2}\left(
-8\gamma ^{\nu }\gamma ^{\beta }-4g^{\nu \beta }\right) .
\end{eqnarray}

Now, plugging $g_{\mu \alpha }$ in the expression,
\begin{eqnarray}
&&g_{\mu \alpha }\left( {\rlap{\hbox{$\mskip
1 mu /$}}}p_{1}+{\rlap{\hbox{$\mskip
1 mu /$}}}q\right) \left( \left\vert A_{2}^{L}\right\vert
^{2}P_{R}+\left\vert A_{2}^{R}\right\vert ^{2}P_{L}\right) \sigma ^{\mu \nu
}\left( {\rlap{\hbox{$\mskip
1 mu /$}}}p_{2}+m_{2}\right) \sigma ^{\alpha \beta } \notag \\
&=&-\frac{1}{8}[\left( {\rlap{\hbox{$\mskip
1 mu /$}}}p_{1}\left\vert A_{2}^{L}\right\vert ^{2}+{%
\rlap{\hbox{$\mskip
1 mu /$}}}p_{1}\left\vert A_{2}^{L}\right\vert ^{2}\gamma _{5}+{%
\rlap{\hbox{$\mskip
1 mu /$}}}p_{1}\left\vert A_{2}^{R}\right\vert ^{2}-{%
\rlap{\hbox{$\mskip
1 mu /$}}}p_{1}\left\vert A_{2}^{R}\right\vert ^{2}\gamma _{5}\right) \notag \\
&&+\left( {\rlap{\hbox{$\mskip
1 mu /$}}}q\left\vert A_{2}^{L}\right\vert ^{2}+{%
\rlap{\hbox{$\mskip
1 mu /$}}}q\left\vert A_{2}^{L}\right\vert ^{2}\gamma _{5}+{%
\rlap{\hbox{$\mskip
1 mu /$}}}q\left\vert A_{2}^{R}\right\vert ^{2}-{%
\rlap{\hbox{$\mskip
1 mu /$}}}q\left\vert A_{2}^{R}\right\vert ^{2}\gamma _{5}\right) ]\times \notag \\
&&\left[ p_{2\eta }\left( 4g^{\nu \eta }\gamma ^{\beta }+4\gamma ^{\nu
}\gamma ^{\eta }\gamma ^{\beta }+4\gamma ^{\nu }g^{\eta \beta }\right)
+m_{2}\left( -8\gamma ^{\nu }\gamma ^{\beta }-4g^{\nu \beta }\right) \right]
, \notag \\
&=&-\frac{1}{8}[\rlap{\hbox{$\mskip
1 mu /$}}p_{1}\left( \left\vert A_{2}^{L}\right\vert
^{2}+\left\vert A_{2}^{R}\right\vert ^{2}\right) +\rlap{\hbox{$\mskip
1 mu /$}}p_{1}\left(
\left\vert A_{2}^{L}\right\vert ^{2}-\left\vert A_{2}^{R}\right\vert
^{2}\right) \gamma _{5} \notag \\
&&+\rlap{\hbox{$\mskip
1 mu /$}}q\left( \left\vert A_{2}^{L}\right\vert ^{2}+\left\vert
A_{2}^{R}\right\vert ^{2}\right) +\rlap{\hbox{$\mskip
1 mu /$}}q\left( \left\vert
A_{2}^{L}\right\vert ^{2}-\left\vert A_{2}^{R}\right\vert ^{2}\right) \gamma
_{5}]\times \notag \\
&&\left[ p_{2\eta }\left( 4g^{\nu \eta }\gamma ^{\beta }+4\gamma ^{\nu
}\gamma ^{\eta }\gamma ^{\beta }+4\gamma ^{\nu }g^{\eta \beta }\right)
+m_{2}\left( -8\gamma ^{\nu }\gamma ^{\beta }-4g^{\nu \beta }\right) \right].
\end{eqnarray}
We can written the equation above as
\begin{eqnarray}
&=&-\frac{1}{8}\{\left[ \rlap{\hbox{$\mskip
1 mu /$}}p_{1}\left( \left\vert A_{2}^{L}\right\vert
^{2}+\left\vert A_{2}^{R}\right\vert ^{2}\right) +\rlap{\hbox{$\mskip
1 mu /$}}p_{1}\left(
\left\vert A_{2}^{L}\right\vert ^{2}-\left\vert A_{2}^{R}\right\vert
^{2}\right) \gamma _{5}\right] \left[ p_{2\eta }\left( 4g^{\nu \eta }\gamma
^{\beta }+4\gamma ^{\nu }\gamma ^{\eta }\gamma ^{\beta }+4\gamma ^{\nu
}g^{\eta \beta }\right) \right] \notag \\
&&+\left[ \rlap{\hbox{$\mskip
1 mu /$}}p_{1}\left( \left\vert A_{2}^{L}\right\vert ^{2}+\left\vert
A_{2}^{R}\right\vert ^{2}\right) +\rlap{\hbox{$\mskip
1 mu /$}}p_{1}\left( \left\vert
A_{2}^{L}\right\vert ^{2}-\left\vert A_{2}^{R}\right\vert ^{2}\right) \gamma
_{5}\right] m_{2}\left( -8\gamma ^{\nu }\gamma ^{\beta }-4g^{\nu \beta
}\right) \notag \\
&&+\left[ \rlap{\hbox{$\mskip
1 mu /$}}q\left( \left\vert A_{2}^{L}\right\vert ^{2}+\left\vert
A_{2}^{R}\right\vert ^{2}\right) +\rlap{\hbox{$\mskip
1 mu /$}}q\left( \left\vert
A_{2}^{L}\right\vert ^{2}-\left\vert A_{2}^{R}\right\vert ^{2}\right) \gamma
_{5}\right] \left[ p_{2\eta }\left( 4g^{\nu \eta }\gamma ^{\beta }+4\gamma
^{\nu }\gamma ^{\eta }\gamma ^{\beta }+4\gamma ^{\nu }g^{\eta \beta }\right) %
\right] \notag \\
&&+\left[ \rlap{\hbox{$\mskip
1 mu /$}}q\left( \left\vert A_{2}^{L}\right\vert ^{2}+\left\vert
A_{2}^{R}\right\vert ^{2}\right) +\rlap{\hbox{$\mskip
1 mu /$}}q\left( \left\vert
A_{2}^{L}\right\vert ^{2}-\left\vert A_{2}^{R}\right\vert ^{2}\right) \gamma
_{5}\right] m_{2}\left( -8\gamma ^{\nu }\gamma ^{\beta }-4g^{\nu \beta
}\right) \}.
\end{eqnarray}
It follows that
\begin{eqnarray}
&=&-\frac{1}{8}\{4\left[ \rlap{\hbox{$\mskip
1 mu /$}}p_{1}\left( \left\vert A_{2}^{L}\right\vert
^{2}+\left\vert A_{2}^{R}\right\vert ^{2}\right) +\rlap{\hbox{$\mskip
1 mu /$}}p_{1}\left(
\left\vert A_{2}^{L}\right\vert ^{2}-\left\vert A_{2}^{R}\right\vert
^{2}\right) \gamma _{5}\right] \left( p_{2}^{\nu }\gamma ^{\beta }+p_{2\eta
}\gamma ^{\nu }\gamma ^{\eta }\gamma ^{\beta }+p_{2}^{\beta }\gamma ^{\nu
}\right) \notag \\
&&-4m_{2}\left[ \rlap{\hbox{$\mskip
1 mu /$}}p_{1}\left( \left\vert A_{2}^{L}\right\vert
^{2}+\left\vert A_{2}^{R}\right\vert ^{2}\right) +\rlap{\hbox{$\mskip
1 mu /$}}p_{1}\left(
\left\vert A_{2}^{L}\right\vert ^{2}-\left\vert A_{2}^{R}\right\vert
^{2}\right) \gamma _{5}\right] \left( 2\gamma ^{\nu }\gamma ^{\beta }+g^{\nu
\beta }\right) \notag \\
&&+\left[ \rlap{\hbox{$\mskip
1 mu /$}}q\left( \left\vert A_{2}^{L}\right\vert ^{2}+\left\vert
A_{2}^{R}\right\vert ^{2}\right) +\rlap{\hbox{$\mskip
1 mu /$}}q\left( \left\vert
A_{2}^{L}\right\vert ^{2}-\left\vert A_{2}^{R}\right\vert ^{2}\right) \gamma
_{5}\right] \left( 4p_{2}^{\nu }\gamma ^{\beta }+4p_{2\eta }\gamma ^{\nu
}\gamma ^{\eta }\gamma ^{\beta }+4p_{2}^{\beta }\gamma ^{\nu }\right) \notag \\
&&-4m_{2}\left[ \rlap{\hbox{$\mskip
1 mu /$}}q\left( \left\vert A_{2}^{L}\right\vert
^{2}+\left\vert A_{2}^{R}\right\vert ^{2}\right) +\rlap{\hbox{$\mskip
1 mu /$}}q\left( \left\vert
A_{2}^{L}\right\vert ^{2}-\left\vert A_{2}^{R}\right\vert ^{2}\right) \gamma
_{5}\right] \left( 2\gamma ^{\nu }\gamma ^{\beta }+g^{\nu \beta }\right) \}.
\end{eqnarray}

Firstly, in order to simplify the evaluation, let's calculate the term only with $\rlap{\hbox{$\mskip 
1 mu /$}}p_{1}$, as follows below
\begin{eqnarray}
&=&-\frac{1}{2}\{\left[ \rlap{\hbox{$\mskip
1 mu /$}}p_{1}\left( \left\vert A_{2}^{L}\right\vert
^{2}+\left\vert A_{2}^{R}\right\vert ^{2}\right) +\rlap{\hbox{$\mskip
1 mu /$}}p_{1}\left(
\left\vert A_{2}^{L}\right\vert ^{2}-\left\vert A_{2}^{R}\right\vert
^{2}\right) \gamma _{5}\right] \left( p_{2}^{\nu }\gamma ^{\beta }+p_{2\eta
}\gamma ^{\nu }\gamma ^{\eta }\gamma ^{\beta }+p_{2}^{\beta }\gamma ^{\nu
}\right)  \notag \\
&&-m_{2}\left[ \rlap{\hbox{$\mskip
1 mu /$}}p_{1}\left( \left\vert A_{2}^{L}\right\vert
^{2}+\left\vert A_{2}^{R}\right\vert ^{2}\right) +\rlap{\hbox{$\mskip
1 mu /$}}p_{1}\left(
\left\vert A_{2}^{L}\right\vert ^{2}-\left\vert A_{2}^{R}\right\vert
^{2}\right) \gamma _{5}\right] \left( 2\gamma ^{\nu }\gamma ^{\beta }+g^{\nu
\beta }\right) , \notag \\
&=&-\frac{1}{2}\{\left( \left\vert A_{2}^{L}\right\vert ^{2}+\left\vert
A_{2}^{R}\right\vert ^{2}\right) \rlap{\hbox{$\mskip
1 mu /$}}p_{1}\left( p_{2}^{\nu }\gamma
^{\beta }+p_{2\eta }\gamma ^{\nu }\gamma ^{\eta }\gamma ^{\beta
}+p_{2}^{\beta }\gamma ^{\nu }\right)  \notag \\
&&+\left( \left\vert A_{2}^{L}\right\vert ^{2}-\left\vert
A_{2}^{R}\right\vert ^{2}\right) \rlap{\hbox{$\mskip
1 mu /$}}p_{1}\gamma _{5}\left( p_{2}^{\nu
}\gamma ^{\beta }+p_{2\eta }\gamma ^{\nu }\gamma ^{\eta }\gamma ^{\beta
}+p_{2}^{\beta }\gamma ^{\nu }\right)  \notag \\
&&-m_{2}\left( \left\vert A_{2}^{L}\right\vert ^{2}+\left\vert
A_{2}^{R}\right\vert ^{2}\right) \rlap{\hbox{$\mskip
1 mu /$}}p_{1}\left( 2\gamma ^{\nu }\gamma
^{\beta }+g^{\nu \beta }\right)  \notag \\
&&-m_{2}\left( \left\vert A_{2}^{L}\right\vert ^{2}-\left\vert
A_{2}^{R}\right\vert ^{2}\right) \rlap{\hbox{$\mskip
1 mu /$}}p_{1}\gamma _{5}\left( 2\gamma ^{\nu
}\gamma ^{\beta }+g^{\nu \beta }\right) \}.
\end{eqnarray}

Let's calculate the term in the braces explicitly%
\begin{eqnarray}
\{...\}&=&\left( \left\vert A_{2}^{L}\right\vert ^{2}+\left\vert
A_{2}^{R}\right\vert ^{2}\right) \rlap{\hbox{$\mskip
1 mu /$}}p_{1}\left( p_{2}^{\nu }\gamma
^{\beta }+p_{2\eta }\gamma ^{\nu }\gamma ^{\eta }\gamma ^{\beta
}+p_{2}^{\beta }\gamma ^{\nu }\right)  \notag \\
&&+\left( \left\vert A_{2}^{L}\right\vert ^{2}-\left\vert
A_{2}^{R}\right\vert ^{2}\right) \rlap{\hbox{$\mskip
1 mu /$}}p_{1}\gamma _{5}\left( p_{2}^{\nu
}\gamma ^{\beta }+p_{2\eta }\gamma ^{\nu }\gamma ^{\eta }\gamma ^{\beta
}+p_{2}^{\beta }\gamma ^{\nu }\right) \notag \\
&&-m_{2}\left( \left\vert A_{2}^{L}\right\vert ^{2}+\left\vert
A_{2}^{R}\right\vert ^{2}\right) \rlap{\hbox{$\mskip
1 mu /$}}p_{1}\left( 2\gamma ^{\nu }\gamma
^{\beta }+g^{\nu \beta }\right)  \notag \\
&&-m_{2}\left( \left\vert A_{2}^{L}\right\vert ^{2}-\left\vert
A_{2}^{R}\right\vert ^{2}\right) \rlap{\hbox{$\mskip
1 mu /$}}p_{1}\gamma _{5}\left( 2\gamma ^{\nu
}\gamma ^{\beta }+g^{\nu \beta }\right) \}.
\end{eqnarray}

With $\rlap{\hbox{$\mskip
1 mu /$}}p_{1}=p_{1\alpha }\gamma ^{\alpha },$ we get%
\begin{eqnarray}
&&\left( \left\vert A_{2}^{L}\right\vert ^{2}+\left\vert
A_{2}^{R}\right\vert ^{2}\right) p_{1\alpha }\left( p_{2}^{\nu }\gamma
^{\alpha }\gamma ^{\beta }+p_{2\eta }\gamma ^{\alpha }\gamma ^{\nu }\gamma
^{\eta }\gamma ^{\beta }+p_{2}^{\beta }\gamma ^{\alpha }\gamma ^{\nu
}\right)  \notag \\
&&+\left( \left\vert A_{2}^{L}\right\vert ^{2}-\left\vert
A_{2}^{R}\right\vert ^{2}\right) p_{1\alpha }\left( p_{2}^{\nu }\gamma
^{\alpha }\gamma _{5}\gamma ^{\beta }+p_{2\eta }\gamma ^{\alpha }\gamma
_{5}\gamma ^{\nu }\gamma ^{\eta }\gamma ^{\beta }+p_{2}^{\beta }\gamma
^{\alpha }\gamma _{5}\gamma ^{\nu }\right)  \notag \\
&&-m_{2}\left( \left\vert A_{2}^{L}\right\vert ^{2}+\left\vert
A_{2}^{R}\right\vert ^{2}\right) p_{1\alpha }\left( 2\gamma ^{\alpha }\gamma
^{\nu }\gamma ^{\beta }+g^{\nu \beta }\gamma ^{\alpha }\right)  \notag \\
&&-m_{2}\left( \left\vert A_{2}^{L}\right\vert ^{2}-\left\vert
A_{2}^{R}\right\vert ^{2}\right) p_{1\alpha }\left( 2\gamma ^{\alpha }\gamma
_{5}\gamma ^{\nu }\gamma ^{\beta }+g^{\nu \beta }\gamma ^{\alpha }\gamma
_{5}\right).
\end{eqnarray}
Putting in evidence $p_{1\alpha }$ in the equation above, it leads to
\begin{eqnarray}
&=&p_{1\alpha }\{\left( \left\vert A_{2}^{L}\right\vert ^{2}+\left\vert
A_{2}^{R}\right\vert ^{2}\right) \left( p_{2}^{\nu }\gamma ^{\alpha }\gamma
^{\beta }+p_{2\eta }\gamma ^{\alpha }\gamma ^{\nu }\gamma ^{\eta }\gamma
^{\beta }+p_{2}^{\beta }\gamma ^{\alpha }\gamma ^{\nu }\right)  \notag \\
&&+\left( \left\vert A_{2}^{L}\right\vert ^{2}-\left\vert
A_{2}^{R}\right\vert ^{2}\right) \left( p_{2}^{\nu }\gamma ^{\alpha }\gamma
_{5}\gamma ^{\beta }+p_{2\eta }\gamma ^{\alpha }\gamma _{5}\gamma ^{\nu
}\gamma ^{\eta }\gamma ^{\beta }+p_{2}^{\beta }\gamma ^{\alpha }\gamma
_{5}\gamma ^{\nu }\right)  \notag \\
&&-m_{2}\left( \left\vert A_{2}^{L}\right\vert ^{2}+\left\vert
A_{2}^{R}\right\vert ^{2}\right) \left( 2\gamma ^{\alpha }\gamma ^{\nu
}\gamma ^{\beta }+g^{\nu \beta }\gamma ^{\alpha }\right)   \notag \\
&&-m_{2}\left( \left\vert A_{2}^{L}\right\vert ^{2}-\left\vert
A_{2}^{R}\right\vert ^{2}\right) \left( 2\gamma ^{\alpha }\gamma _{5}\gamma
^{\nu }\gamma ^{\beta }+g^{\nu \beta }\gamma ^{\alpha }\gamma _{5}\right) \}.
\end{eqnarray}

Taking the trace, we have%
\begin{eqnarray}
&=&p_{1\alpha }\{\left( \left\vert A_{2}^{L}\right\vert ^{2}+\left\vert
A_{2}^{R}\right\vert ^{2}\right) \left[ p_{2}^{\nu }\text{Tr}\left( \gamma ^{\alpha
}\gamma ^{\beta }\right) +p_{2\eta }\text{Tr}\left( \gamma ^{\alpha }\gamma ^{\nu
}\gamma ^{\eta }\gamma ^{\beta }\right) +p_{2}^{\beta }\text{Tr}\left( \gamma
^{\alpha }\gamma ^{\nu }\right) \right]  \notag \\
&&+\left( \left\vert A_{2}^{L}\right\vert ^{2}-\left\vert
A_{2}^{R}\right\vert ^{2}\right) \left[ p_{2}^{\nu }\text{Tr}\left( \gamma ^{\alpha
}\gamma _{5}\gamma ^{\beta }\right) +p_{2\eta }\text{Tr}\left( \gamma ^{\alpha
}\gamma _{5}\gamma ^{\nu }\gamma ^{\eta }\gamma ^{\beta }\right)
+p_{2}^{\beta }\text{Tr}\left( \gamma ^{\alpha }\gamma _{5}\gamma ^{\nu }\right) %
\right]  \notag \\
&&-m_{2}\left( \left\vert A_{2}^{L}\right\vert ^{2}+\left\vert
A_{2}^{R}\right\vert ^{2}\right) \left[ 2\text{Tr}\left( \gamma ^{\alpha }\gamma
^{\nu }\gamma ^{\beta }\right) \right]-m_{2}\left( \left\vert A_{2}^{L}\right\vert ^{2}-\left\vert
A_{2}^{R}\right\vert ^{2}\right) \left[ 2\text{Tr}\left( \gamma ^{\alpha }\gamma
_{5}\gamma ^{\nu }\gamma ^{\beta }\right) \right] \}. \notag \\
\end{eqnarray}

Observing that the trace of an odd product of $\gamma _{\nu }$vanishes and using the trace identities,%
\begin{eqnarray}
\text{Tr}\left( \gamma ^{\mu }\gamma ^{\nu }\right)  &=&4g^{\mu \nu }, \\
\text{Tr}(\gamma ^{\mu }\gamma ^{\nu }\gamma _{5}) &=&0, \\
\text{Tr}(\gamma ^{\mu }\gamma ^{\nu }\gamma ^{\alpha }\gamma ^{\beta }) &=&4\left(
g^{\mu \nu }g^{\alpha \beta }-g^{\mu \alpha }g^{\nu \beta }+g^{\mu \beta
}g^{\nu \alpha }\right) , \\
\text{Tr}(\gamma _{5}\gamma ^{\mu }\gamma ^{\nu }\gamma ^{\alpha }\gamma ^{\beta })
&=&-4i\varepsilon ^{\mu \nu \eta \beta },
\end{eqnarray}%
we get the result,%
\begin{eqnarray}
&=&p_{1\alpha }\{\left( \left\vert A_{2}^{L}\right\vert ^{2}+\left\vert
A_{2}^{R}\right\vert ^{2}\right) \left[ 4p_{2}^{\nu }g^{\alpha \beta
}+4p_{2\eta }\left( g^{\alpha \nu }g^{\eta \beta }-g^{\alpha \eta }g^{\nu
\beta }+g^{\alpha \beta }g^{\nu \eta }\right) +4p_{2}^{\beta }g^{\nu \alpha }%
\right]   \notag \\
&&+\left( \left\vert A_{2}^{L}\right\vert ^{2}-\left\vert
A_{2}^{R}\right\vert ^{2}\right) \left[ -4ip_{2\eta }\varepsilon ^{\nu \eta
\beta \alpha }\right] \}.
\end{eqnarray}

Inserting the\ product of photon 4-momenta $q_{\nu }q_{\beta },$ we have%
\begin{eqnarray}
&&q_{\nu }q_{\beta }p_{1\alpha }\{\left( \left\vert A_{2}^{L}\right\vert
^{2}+\left\vert A_{2}^{R}\right\vert ^{2}\right) \left[ 4p_{2}^{\nu
}g^{\alpha \beta }+4p_{2\eta }\left( g^{\alpha \nu }g^{\eta \beta
}-g^{\alpha \eta }g^{\nu \beta }+g^{\alpha \beta }g^{\nu \eta }\right)
+4p_{2}^{\beta }g^{\nu \alpha }\right]   \notag \\
&&+\left( \left\vert A_{2}^{L}\right\vert ^{2}-\left\vert
A_{2}^{R}\right\vert ^{2}\right) \left[ -4ip_{2\eta }\varepsilon ^{\nu \eta
\beta \alpha }\right] \}  \notag \\
&=&4\{\left( \left\vert A_{2}^{L}\right\vert ^{2}+\left\vert
A_{2}^{R}\right\vert ^{2}\right) [q_{\nu }q_{\beta }p_{1\alpha }p_{2}^{\nu
}g^{\alpha \beta }  \notag \\
&&+\left( q_{\nu }q_{\beta }p_{1\alpha }p_{2\eta }g^{\alpha \nu }g^{\eta
\beta }-q_{\nu }q_{\beta }p_{1\alpha }p_{2\eta }g^{\alpha \eta }g^{\nu \beta
}+q_{\nu }q_{\beta }p_{1\alpha }p_{2\eta }g^{\alpha \beta }g^{\nu \eta
}\right) +q_{\nu }q_{\beta }p_{1\alpha }p_{2}^{\beta }g^{\nu \alpha }]
\notag \\
&&+\left( \left\vert A_{2}^{L}\right\vert ^{2}-\left\vert
A_{2}^{R}\right\vert ^{2}\right) \left[ -4iq_{\nu }q_{\beta }p_{1\alpha
}p_{2\eta }\varepsilon ^{\nu \eta \beta \alpha }\right]   .
\end{eqnarray}

Taking the contractions in the equation above, we get%
\begin{eqnarray}
&=&4\{\left( \left\vert A_{2}^{L}\right\vert ^{2}+\left\vert
A_{2}^{R}\right\vert ^{2}\right) [\left( p_{1}.q\right) \left(
p_{2}.q\right)
+\left( \left( p_{1}.q\right) \left( p_{2}.q\right) -\left(
p_{1}.p_{2}\right) \left( q.q\right) +\left( p_{1}.q\right) \left(
p_{2}.q\right) \right) +\left( p_{1}.q\right) \left( p_{2}.q\right) ]  \notag
\\
&&+\left( \left\vert A_{2}^{L}\right\vert ^{2}-\left\vert
A_{2}^{R}\right\vert ^{2}\right) \left[ -4iq_{\nu }q_{\beta }p_{1\alpha
}p_{2\eta }\varepsilon ^{\nu \eta \beta \alpha }\right] \},  \notag \\
&=&\{16\left( \left\vert A_{2}^{L}\right\vert ^{2}+\left\vert
A_{2}^{R}\right\vert ^{2}\right) \left[ \left( p_{1}.q\right) \left(
p_{2}.q\right) \right]-4\left( \left\vert A_{2}^{L}\right\vert ^{2}-\left\vert
A_{2}^{R}\right\vert ^{2}\right) \left[ i\varepsilon ^{\nu \eta \beta \alpha
}q_{\nu }q_{\beta }p_{1\alpha }p_{2\eta }\right] \}.
\end{eqnarray}

Given that%
\begin{eqnarray*}
\varepsilon ^{\nu \eta \beta \alpha }q_{\nu }q_{\beta }p_{1\alpha }p_{2\eta
} &=&\varepsilon ^{\beta \nu \eta \alpha }q_{\nu }q_{\beta }p_{1\alpha
}p_{2\eta }=\varepsilon ^{\eta \nu \beta \alpha }q_{\beta }q_{\nu
}p_{1\alpha }p_{2\eta }=-\varepsilon ^{\nu \eta \beta \alpha }q_{\nu
}q_{\beta }p_{1\alpha }p_{2\eta } \\
2\varepsilon ^{\nu \eta \beta \alpha }q_{\nu }q_{\beta }p_{1\alpha }p_{2\eta
} &=&0,
\end{eqnarray*}
it follows,
\begin{eqnarray}
&&q_{\nu }q_{\beta }p_{1\alpha }\{\left( \left\vert A_{2}^{L}\right\vert
^{2}+\left\vert A_{2}^{R}\right\vert ^{2}\right) \left[ 4p_{2}^{\nu
}g^{\alpha \beta }+4p_{2\eta }\left( g^{\alpha \nu }g^{\eta \beta
}-g^{\alpha \eta }g^{\nu \beta }+g^{\alpha \beta }g^{\nu \eta }\right)
+4p_{2}^{\beta }g^{\nu \alpha }\right]   \notag \\
&&+\left( \left\vert A_{2}^{L}\right\vert ^{2}-\left\vert
A_{2}^{R}\right\vert ^{2}\right) \left[ -4p_{2\eta }\varepsilon ^{\nu \eta
\beta \alpha }\right] \}  \notag \\
&=&16\left( \left\vert A_{2}^{L}\right\vert ^{2}+\left\vert
A_{2}^{R}\right\vert ^{2}\right) \left[ \left( p_{1}.q\right) \left(
p_{2}.q\right) \right].
\end{eqnarray}

Analogously to what we calculated above, now we will proceed onto the evaluation of the term with $\rlap{\hbox{$\mskip
1 mu /$}}q$:%
\begin{eqnarray}
&&\left[ \rlap{\hbox{$\mskip
1 mu /$}}q\left( \left\vert A_{2}^{L}\right\vert ^{2}+\left\vert
A_{2}^{R}\right\vert ^{2}\right) +\rlap{\hbox{$\mskip
1 mu /$}}q\left( \left\vert
A_{2}^{L}\right\vert ^{2}-\left\vert A_{2}^{R}\right\vert ^{2}\right) \gamma
_{5}\right] \left( p_{2}^{\nu }\gamma ^{\beta }+p_{2\eta }\gamma ^{\nu
}\gamma ^{\eta }\gamma ^{\beta }+p_{2}^{\beta }\gamma ^{\nu }\right)   \notag
\\
&&-m_{2}\left[ \rlap{\hbox{$\mskip
1 mu /$}}q\left( \left\vert A_{2}^{L}\right\vert ^{2}+\left\vert
A_{2}^{R}\right\vert ^{2}\right) +\rlap{\hbox{$\mskip
1 mu /$}}q\left( \left\vert
A_{2}^{L}\right\vert ^{2}-\left\vert A_{2}^{R}\right\vert ^{2}\right) \gamma
_{5}\right] \left( 2\gamma ^{\nu }\gamma ^{\beta }+g^{\nu \beta }\right),
\notag \\
&=&-\frac{1}{2}\{\left( \left\vert A_{2}^{L}\right\vert ^{2}+\left\vert
A_{2}^{R}\right\vert ^{2}\right) \rlap{\hbox{$\mskip
1 mu /$}}q\left( p_{2}^{\nu }\gamma ^{\beta
}+p_{2\eta }\gamma ^{\nu }\gamma ^{\eta }\gamma ^{\beta }+p_{2}^{\beta
}\gamma ^{\nu }\right)   \notag \\
&&+\left( \left\vert A_{2}^{L}\right\vert ^{2}-\left\vert
A_{2}^{R}\right\vert ^{2}\right) \rlap{\hbox{$\mskip
1 mu /$}}q\gamma _{5}\left( p_{2}^{\nu }\gamma
^{\beta }+p_{2\eta }\gamma ^{\nu }\gamma ^{\eta }\gamma ^{\beta
}+p_{2}^{\beta }\gamma ^{\nu }\right)   \notag \\
&&-m_{2}\left( \left\vert A_{2}^{L}\right\vert ^{2}+\left\vert
A_{2}^{R}\right\vert ^{2}\right) \rlap{\hbox{$\mskip
1 mu /$}}q\left( 2\gamma ^{\nu }\gamma ^{\beta
}+g^{\nu \beta }\right)   \notag \\
&&-m_{2}\left( \left\vert A_{2}^{L}\right\vert ^{2}-\left\vert
A_{2}^{R}\right\vert ^{2}\right) \rlap{\hbox{$\mskip
1 mu /$}}q\gamma _{5}\left( 2\gamma ^{\nu
}\gamma ^{\beta }+g^{\nu \beta }\right) \}.
\end{eqnarray}

With $\rlap{\hbox{$\mskip
1 mu /$}}q=q_{\mu }\gamma ^{\mu },$ we get
\begin{eqnarray}
&=&q_{\alpha }\{\left( \left\vert A_{2}^{L}\right\vert ^{2}+\left\vert
A_{2}^{R}\right\vert ^{2}\right) \left( p_{2}^{\nu }\gamma ^{\alpha }\gamma
^{\beta }+p_{2\eta }\gamma ^{\alpha }\gamma ^{\nu }\gamma ^{\eta }\gamma
^{\beta }+p_{2}^{\beta }\gamma ^{\alpha }\gamma ^{\nu }\right)   \notag \\
&&+\left( \left\vert A_{2}^{L}\right\vert ^{2}-\left\vert
A_{2}^{R}\right\vert ^{2}\right) \left( p_{2}^{\nu }\gamma ^{\alpha }\gamma
_{5}\gamma ^{\beta }+p_{2\eta }\gamma ^{\alpha }\gamma _{5}\gamma ^{\nu
}\gamma ^{\eta }\gamma ^{\beta }+p_{2}^{\beta }\gamma ^{\alpha }\gamma
_{5}\gamma ^{\nu }\right)   \notag \\
&&-m_{2}\left( \left\vert A_{2}^{L}\right\vert ^{2}+\left\vert
A_{2}^{R}\right\vert ^{2}\right) \left( 2\gamma ^{\alpha }\gamma ^{\nu
}\gamma ^{\beta }+g^{\nu \beta }\gamma ^{\alpha }\right)   \notag \\
&&-m_{2}\left( \left\vert A_{2}^{L}\right\vert ^{2}-\left\vert
A_{2}^{R}\right\vert ^{2}\right) \left( 2\gamma ^{\alpha }\gamma _{5}\gamma
^{\nu }\gamma ^{\beta }+g^{\nu \beta }\gamma ^{\alpha }\gamma _{5}\right) \}.
\end{eqnarray}

Taking the trace of the equation above%
\begin{eqnarray}
&=&q_{\alpha }\{\left( \left\vert A_{2}^{L}\right\vert ^{2}+\left\vert
A_{2}^{R}\right\vert ^{2}\right) \left[ p^{2\nu }\text{Tr}\left( \gamma ^{\alpha
}\gamma ^{\beta }\right) +p_{2\eta }\text{Tr}\left( \gamma ^{\alpha }\gamma ^{\nu
}\gamma ^{\eta }\gamma ^{\beta }\right) +p_{2}^{\beta }\text{Tr}\left( \gamma
^{\alpha }\gamma ^{\nu }\right) \right]   \notag \\
&&+\left( \left\vert A_{2}^{L}\right\vert ^{2}-\left\vert
A_{2}^{R}\right\vert ^{2}\right) \left[ p^{2\nu }\text{Tr}\left( \gamma ^{\alpha
}\gamma _{5}\gamma ^{\beta }\right) +p_{2\eta }\text{Tr}\left( \gamma ^{\alpha
}\gamma _{5}\gamma ^{\nu }\gamma ^{\eta }\gamma ^{\beta }\right)
+p_{2}^{\beta }\text{Tr}\left( \gamma ^{\alpha }\gamma _{5}\gamma ^{\nu }\right) %
\right]   \notag \\
&&-m_{2}\left( \left\vert A_{2}^{L}\right\vert ^{2}+\left\vert
A_{2}^{R}\right\vert ^{2}\right) \left[ 2\text{Tr}\left( \gamma ^{\alpha }\gamma
^{\nu }\gamma ^{\beta }\right) \right]   
-m_{2}\left( \left\vert A_{2}^{L}\right\vert ^{2}-\left\vert
A_{2}^{R}\right\vert ^{2}\right) \left[ 2\text{Tr}\left( \gamma ^{\alpha }\gamma
_{5}\gamma ^{\nu }\gamma ^{\beta }\right) \right] \}, \notag
\end{eqnarray}%
we get%
\begin{eqnarray}
&=&q_{\alpha }\{\left( \left\vert A_{2}^{L}\right\vert ^{2}+\left\vert
A_{2}^{R}\right\vert ^{2}\right) \left[ p^{2\nu }\text{Tr}\left( \gamma ^{\alpha
}\gamma ^{\beta }\right) +p_{2\eta }\text{Tr}\left( \gamma ^{\alpha }\gamma ^{\nu
}\gamma ^{\eta }\gamma ^{\beta }\right) +p_{2}^{\beta }\text{Tr}\left( \gamma
^{\alpha }\gamma ^{\nu }\right) \right]   \notag \\
&&+\left( \left\vert A_{2}^{L}\right\vert ^{2}-\left\vert
A_{2}^{R}\right\vert ^{2}\right) \left[ p_{2\eta }\text{Tr}\left( \gamma
^{\alpha }\gamma _{5}\gamma ^{\nu }\gamma ^{\eta }\gamma ^{\beta }\right) %
\right] \}.
\end{eqnarray}%
Using the trace identities, we have
\begin{eqnarray}
&=&q_{\alpha }\{\left( \left\vert A_{2}^{L}\right\vert ^{2}+\left\vert
A_{2}^{R}\right\vert ^{2}\right) \left[ 4p^{2\nu }g^{\alpha \beta
}+4p_{2\eta }\left( g^{\alpha \nu }g^{\eta \beta }-g^{\alpha \eta }g^{\nu
\beta }+g^{\alpha \beta }g^{\nu \eta }\right) +4p_{2}^{\beta }g^{\nu \alpha }%
\right]   \notag \\
&&+\left( \left\vert A_{2}^{L}\right\vert ^{2}-\left\vert
A_{2}^{R}\right\vert ^{2}\right) \left[ -4p_{2\eta }\varepsilon ^{\nu \eta
\beta \alpha }\right] \}.
\end{eqnarray}

Plugging the $q_{\nu }q_{\beta }$ in the equation above, we get
\begin{eqnarray}
&&q_{\nu }q_{\beta }q_{\alpha }\{\left( \left\vert A_{2}^{L}\right\vert
^{2}+\left\vert A_{2}^{R}\right\vert ^{2}\right) \left[ 4p^{2\nu }g^{\alpha
\beta }+4p_{2\eta }\left( g^{\alpha \nu }g^{\eta \beta }-g^{\alpha \eta
}g^{\nu \beta }+g^{\alpha \beta }g^{\nu \eta }\right) +4p_{2}^{\beta }g^{\nu
\alpha }\right]   \notag \\
&&+\left( \left\vert A_{2}^{L}\right\vert ^{2}-\left\vert
A_{2}^{R}\right\vert ^{2}\right) \left[ -4p_{2\eta }\varepsilon ^{\nu \eta
\beta \alpha }\right] \},  \notag \\
&=&4\left( \left\vert A_{2}^{L}\right\vert ^{2}+\left\vert
A_{2}^{R}\right\vert ^{2}\right) (q_{\nu }q_{\beta }q_{\alpha }p^{2\nu
}g^{\alpha \beta }+ q_{\nu }q_{\beta }q_{\alpha }p_{2\eta }g^{\alpha \nu }g^{\eta
\beta }-q_{\nu }q_{\beta }q_{\alpha }p_{2\eta }g^{\alpha \eta }g^{\nu \beta}  \notag \\
&&+q_{\nu }q_{\beta }q_{\alpha }p_{2\eta }g^{\alpha \beta }g^{\nu \eta}
+q_{\nu }q_{\beta }q_{\alpha }p_{2}^{\beta }g^{\nu \alpha })  
+\left( \left\vert A_{2}^{L}\right\vert ^{2}-\left\vert
A_{2}^{R}\right\vert ^{2}\right) \left( -4q_{\nu }q_{\beta }q_{\alpha
}p_{2\eta }\varepsilon ^{\nu \eta \beta \alpha }\right), \notag \\
&=&4\left( \left\vert A_{2}^{L}\right\vert ^{2}+\left\vert
A_{2}^{R}\right\vert ^{2}\right) [\left( p_{2}.q\right) \left( q.q\right)
+ \left( p_{2}.q\right) \left( q.q\right) -\left( p_{2}.q\right)
\left( q.q\right) +\left( q.q\right) \left( p_{2}.q\right) +\left(
q.q\right) \left( p_{2}.q\right) ] \notag \\
&&+\left( \left\vert A_{2}^{L}\right\vert ^{2}-\left\vert
A_{2}^{R}\right\vert ^{2}\right) \left( -4q_{\nu }q_{\beta }q_{\alpha}p_{2\eta }\varepsilon ^{\nu \eta \beta \alpha } \right),
\end{eqnarray}%
where $q^{\mu }q_{\mu }=q.q=q^{2}=0$ (for the case of on-shell photon).

Having in mind the results we obtain the follwing amplitude,
\begin{eqnarray}
\label{amplitude_20}
\overline{\left\vert \mathcal{M}\right\vert ^{2}}&=&\frac{1}{2}%
\sum\limits_{spins}\sum\limits_{\lambda }\left\vert \mathcal{M}\right\vert
^{2}, \notag \\
&=&\frac{1}{2}\frac{1}{2}e^{2}m_{\mu }^{2}\times 16\left( \left\vert
A_{2}^{L}\right\vert ^{2}+\left\vert A_{2}^{R}\right\vert ^{2}\right) \left[
\left( p_{1}.q\right) \left( p_{2}.q\right) \right],
\end{eqnarray}%
where $m_{l_{j}}=m_{\mu }$ (It's the mass of the decaying particle, i.e., the muon mass). This result is essential for the obtaining of the decay rate, since it depends on the spin-averaged squared amplitude.  

In the centre of the mass of the system, the decaying particle is
at rest, then $E=m_{2}=m_{\mu }$ and $\text{p}_{\mu }=0$ $\left( \text{3-momentum
of the muon}\right) .$ The 3-momenta of the final state particles are $%
\text{p}_{\gamma }=-\text{p}^{\ast }$ and $\text{p}_{e}=\text{p}^{\ast }.$ The 4-momentum of the $\mu
^{-},$ $e^{-}$ and $\gamma $ are respectively,%
\begin{equation}
p_{\mu }=\left( m_{\mu },0,0,0\right) ,\text{ }p_{e}=\left(
E_{e},0,0,-\text{p}^{\ast }\right) ,\text{ }p_{\gamma }=q=\left( E_{\gamma
},0,0,\text{p}^{\ast }\right) .
\end{equation}
Here we are using the notation $p^{\alpha}=p $ (for the contravariant 4-vector) .

Taking $E_{e}=\text{p}^{\ast }$ and $E_{\gamma }=\text{p}^{\ast },$ we can written the 4-momenta relatives to the centre of mass frame as:%
\begin{equation}
p_{\mu }=p_{2}=\left( m_{\mu },0,0,0\right) ,\text{ }p_{e}=p_{1}=\left(
\text{p}^{\ast },0,0,-\text{p}^{\ast }\right) ,\text{ }p_{\gamma }=q=\left( \text{p}^{\ast
},0,0,\text{p}^{\ast }\right) .
\end{equation}

The magnitude of 3-momentum of the final-state particles (in the centre of mass frame) is%
\begin{equation}
\label{p_star}
\text{p}^{\ast }=\frac{1}{2m_{\mu }}\sqrt{\left( m_{\mu }^{2}-m_{e}^{2}\right) ^{2}}%
\approx \frac{1}{2}m_{\mu },
\end{equation}%
where we consider $m_{\mu }\gg m_{e}$, i.e., we neglect the electron mass in relation to the muon mass.

In order to write the squared spin-averaged amplitude in a more simplified way, we can write the product of two contractions as,%
\begin{eqnarray}
\left( p_{1}.q\right) \left( p_{2}.q\right)  &=&\frac{1}{2}m_{\mu
}^{2}\times \frac{1}{2}m_{\mu }^{2},  \notag \\
&=&\frac{1}{4}m_{\mu }^{4}.
\end{eqnarray}

Substituting the result above in the Eq. (\ref{amplitude_20}), the squared amplitude (averaged over particle spin) is written as
\begin{eqnarray}
\label{amplitude_final}
\overline{\left\vert \mathcal{M}\right\vert ^{2}} &=&\frac{1}{2}%
\sum\limits_{spins}\sum\limits_{\lambda }\left\vert \mathcal{M}\right\vert
^{2}=\frac{1}{2}e^{2}m_{\mu }^{2}\frac{1}{2}\times 16\left( \left\vert
A_{2}^{L}\right\vert ^{2}+\left\vert A_{2}^{R}\right\vert ^{2}\right) \frac{1%
}{4}m_{\mu }^{4},  \notag \\
&=&e^{2}m_{\mu }^{6}\left( \left\vert A_{2}^{L}\right\vert ^{2}+\left\vert
A_{2}^{R}\right\vert ^{2}\right) .
\end{eqnarray}

Having in hands the squared amplitude, we can obtain the decay rate of the charged lepton flavor violating process for the muon decay in electron plus one photon. For this, we can need to write the decay rate for the two-bodies decay, which is given by%
\begin{eqnarray}
\Gamma _{\mu \rightarrow e\gamma } &=&\frac{\text{p}^{\ast }}{32\pi ^{2}m_{\mu }^{2}%
}\int \overline{\left\vert \mathcal{M}\right\vert ^{2}}d\Omega .
\end{eqnarray}%
Using Eq. (\ref{p_star}) for $p^{*}$, the decay rate is%
\begin{eqnarray}
\Gamma _{\mu \rightarrow e\gamma } &=&
\frac{1}{64\pi ^{2}m_{\mu }}\int \overline{\left\vert \mathcal{M}%
\right\vert ^{2}}d\Omega .
\end{eqnarray}
Now substituting the squared spin-averaged amplitude given in the Eq. (\ref{amplitude_final}), we obtain%
\begin{eqnarray}
\Gamma _{\mu \rightarrow e\gamma } &=&\frac{1}{64\pi ^{2}m_{\mu }}%
e^{2}m_{\mu }^{6}\left( \left\vert A_{2}^{L}\right\vert ^{2}+\left\vert
A_{2}^{R}\right\vert ^{2}\right) \int d\Omega ,  \notag \\
&=&\frac{1}{64\pi ^{2}}e^{2}m_{\mu }^{5}\left( \left\vert
A_{2}^{L}\right\vert ^{2}+\left\vert A_{2}^{R}\right\vert ^{2}\right) 4\pi ,
\notag \\
&=&\frac{1}{16\pi }e^{2}m_{\mu }^{5}\left( \left\vert A_{2}^{L}\right\vert
^{2}+\left\vert A_{2}^{R}\right\vert ^{2}\right) .
\end{eqnarray}

Therefore, the decay rate of the charged lepton flavor violating decay of the
muon is%
\begin{equation}
\Gamma _{\mu \rightarrow e\gamma }=\frac{e^{2}m_{\mu }^{5}}{16\pi }\left(
\left\vert A_{2}^{L}\right\vert ^{2}+\left\vert A_{2}^{R}\right\vert
^{2}\right) .
\end{equation}%
Observe above that the rate decay of charged lepton flavor violating muon
depends on the mass muon, the electric charge and it is proportional to the product of sum of the squares of  $A_{2}^{L}$ and $A_{2}^{R}$.

It's well known that the decay rate for the muon decay $\left( \mu \rightarrow e\overline{\nu }%
_{e}\nu _{\mu }\right) $, in the SM, is%
\begin{equation}
\Gamma _{\mu \rightarrow e\overline{\nu }_{e}\nu _{\mu }}=\frac{%
G_{F}^{2}m_{\mu }^{5}}{192\pi ^{3}},
\end{equation}%
where $G_{F}$ is Fermi constant and $m_{\mu }$ is the muon's mass. This rate decay of the muon does not violate the charged lepton flavor. 

The branching ratio for the $\left( \mu \rightarrow e\overline{\nu }_{e}\nu
_{\mu }\right) $ charged lepton flavor violating process is%
\begin{eqnarray}
BR\left( \mu \rightarrow e\gamma \right)&=&\frac{\Gamma _{\mu \rightarrow
e\gamma }}{\Gamma _{\mu \rightarrow e\overline{\nu }_{e}\nu _{\mu }}} , \notag
\\
&=&\frac{\frac{e^{2}m_{\mu }^{5}}{16\pi }\left( \left\vert
A_{2}^{L}\right\vert ^{2}+\left\vert A_{2}^{R}\right\vert ^{2}\right) }{%
\frac{G_{F}^{2}m_{\mu }^{5}}{192\pi ^{3}}}, \\
&=&\frac{192e^{2}\pi ^{2}}{16G_{F}^{2}}\left( \left\vert
A_{2}^{L}\right\vert ^{2}+\left\vert A_{2}^{R}\right\vert ^{2}\right) .
\end{eqnarray}

Thus, given that $\alpha =e^{2}/4\pi $ (in natural units), we have finally the branching ratio of the
charged lepton flavor violating muon decay for the ($\mu \rightarrow e\gamma
)$ process%
\begin{equation}
\label{branching_final}
BR\left( \mu \rightarrow e\gamma \right) =\frac{3\alpha ^{2}\left( 4\pi
\right) ^{3}}{4G_{F}^{2}}\left( \left\vert A_{2}^{L}\right\vert
^{2}+\left\vert A_{2}^{R}\right\vert ^{2}\right) ,
\end{equation}%
where $\alpha $ is the electromagnetic fine-structure constant, $G_{F}$ is
the Fermi constant. Besides to $\mu \rightarrow e\gamma$ decay, this branching ratio can be used for others decay processes, as $\tau \rightarrow \mu \gamma$.

It's important to point that the branching ratio, given by the Eq. (\ref{branching_final}), is very general, since we can use it for various models, for instance: supersymmetric and $3-3-1$ models. In fact, we need to obtain only $A_{2}^{L,R}$ for each model with its specific contributions related on the particle content.

\bibliographystyle{JHEPfixed}
\bibliography{literature331}
\end{document}